\documentclass[a4paper,fleqn,usenatbib]{mnras}

\usepackage{newtxtext,newtxmath}

\usepackage[T1]{fontenc}
\usepackage{ae,aecompl}

\usepackage{multirow}
\usepackage{wasysym}
\usepackage{mathtools}
\usepackage{comment}

\usepackage{graphicx}	
\usepackage{amsmath}	
\usepackage{amssymb}	

\usepackage{array}

\newcolumntype{x}[1]{>{\centering\arraybackslash\hspace{0pt}}p{#1}}

\title[Spatial structure of several DIB carriers]{Spatial structure of several diffuse interstellar band carriers}

\author[J. Kos]{
Janez Kos,$^{1}$
\\
$^{1}$Sydney Institute for Astronomy, School of Physics, A28, The University of Sydney, NSW 2006, Australia
}

\date{Accepted XXX. Received YYY; in original form ZZZ}

\pubyear{2017}

\begin{document}
\label{firstpage}
\pagerange{\pageref{firstpage}--\pageref{lastpage}}
\maketitle

\begin{abstract}
Diffuse interstellar bands (DIBs) hold a lot of information about the state and the structure of the ISM. Structure can most directly be observed by extensive spectroscopic surveys, including surveys of stars where DIBs are especially important, as they are conveniently found in all observed bands. Large surveys lack the quality of spectra to detect weak DIBs, so  many spectra from  small regions on the sky have to be combined before a sufficient signal-to-noise ratio (SNR) is achieved. However, the clumpiness of the DIB clouds is unknown, which poses a problem, as the measured properties can end up being averaged over a too large area. We use a technique called Gaussian processes to accurately measure profiles of interstellar absorption lines in 145 high SNR and high resolution spectra of hot stars. Together with Bayesian MCMC approach we also get reliable estimates of the uncertainties. We derive scales at which column densities of 18 DIBs, CH, CH$^+$, Ca~I, and Ca~II show some spatial correlation. This correlation scale is associated with the size of the ISM clouds. Scales expressed as the angle on the sky vary significantly from DIB to DIB between $\sim0.23^\circ$ for the DIB at 5512~{\AA} and 3.5$^\circ$ for the DIB at  6196~{\AA}, suggesting that different DIB carriers have different clumpiness but occupy the same general space. Our study includes lines-of-sight all over the northern Milky Way, as well as out of the Galactic plane, covering regions with different physical conditions. The derived correlation scales therefore represent a general image of the Galactic ISM on the scales of $\sim5$~pc to $100$~pc.
\end{abstract}




\section{Introduction}

Diffuse interstellar bands (DIBs) are yet to be understood absorption features observed in VIS and NIR spectra of reddened stars. There are hundreds of known DIBs all sharing some common properties: they are resolved (narrowest ones have a full width at half maximum (FWHM) of around 0.5~{\AA}), they all correlate at least vaguely with the amount of interstellar extinction in the line-of-sight (LOS), and almost none have a confirmed carrier or origin. DIBs were discovered in the early 1920s \citep{heger22} and their interstellar origin has been confirmed soon afterwards \citep{merrill36}. Despite being among the first molecules observed in the ISM, most of the carriers are still not identified. A bright exception are a few DIBs in the NIR, for which the C$_{60}^+$ has been confirmed as a carrier by laboratory studies \citep{campbell15, walker15}. None of the DIBs addressed in this study have confirmed carriers.  Among the proposed carriers are polycyclic aromatic hydrocarbons \citep{leger85, zwet85, salama99}, fullerenes \citep{kroto88, campbell15}, and other hydrocarbons \citep{motylewski00, maier11, krelowski10} from simple molecules with only 4 atoms to long carbon chains. See reviews \citet{krelowski89}, \citet{herbig95}, and \citet{sarre06} for a thorough overview of the field.

It has been shown \citep{krelowski14, baron15} that there must be more than just a few molecules producing DIBs. Most of the strongest DIBs are believed to each belong to a different carrier. Correlation between DIB strengths and correlation with the interstellar reddening is most commonly used to identify DIB families and relations between them \citep{krelowski87, westerlund89, cami97,kos13}. Correlation is not the only relation between DIBs we can observe. Less common are studies of radial velocities and spatial correlations. Most direct approach to this are measurements of three-dimensional structure in large spectroscopic surveys \citep{kos14,zasowski15}. A huge number of observed LOS is the minimal requirement for such approach. Despite many observed LOS the spatial resolution of such maps has been poor.

Studies of DIBs with a fine (<100~pc) spatial resolution are rare, because in order to observe differential strengths of DIBs high quality spectra are needed. Large spectroscopic surveys of stars provide the needed spatial resolution, but rarely include enough spectra with high SNR. Therefore dedicated observations must be made like \citet{vos11}, \citet{loon09}, or \citet{loon13}. All three studies show a highly structured DIBs-bearing ISM in the Galactic plane \citet{vos11} and well away from it \citep{loon09, loon13}. Only a few strongest DIBs were studied in mentioned papers in small regions of the Galaxy (Upper Scorpius, $\omega$ Cen and the Tarantula nebula). The most detailed study of small scale variations is \citep{cordiner13}. They studied DIBs and some other ISM lines in very high SNR spectra of 4 stars in $\rho$~Oph and found variations of $\sim$3\% (expressed the same way as variations between LOS in this paper) in LOS as close as 2.8 arc seconds. It is hard to generalize properties of the ISM observed in these regions over the whole Galaxy. We aim to change this by observing stars over the whole Galactic plane. Poorly sampled but high quality data can be used to statistically constrain the nature of the ISM structure. Instead of producing localized maps, we study small scale variations of DIB strengths through pairwise correlations in tightly separated LOS. 

We can hypothesize three different cases for the structure of the DIB bearing ISM: clouds with different clumpiness for different DIB species, separate clouds of different DIB species, or gradients between different species in a single cloud. From the pairwise correlation between DIB profiles in tight pairs of LOS we can distinguish between the three cases. We can put some constraints on respective carriers from observed differences in spatial distribution between different DIB species. 

We acquired high SNR and medium-high resolution echelle spectra of 145 hot stars covering a wavelength range between 3700 and 7300~{\AA}. Target selection and reduction of the dataset is described in Section~\ref{sec:obs}. Only strong and prominent DIBs are studied in this work. To describe their profiles we fit them with asymmetric Gaussians, empowering the Gaussian processes and a Bayesian MCMC fitting scheme to calculate uncertainties and deal with stellar continuum and statistical noise. The method is described in Section~\ref{sec:method}. The core of this paper is the study of close pairs of LOS, the differences between them and how they reveal the structure of the ISM and physics of the DIBs. This is presented in Section~\ref{sec:pairs}. A discussion of the results and the implications of this study are presented in Section~\ref{sec:disc}.

\section{Observations and reduction}
\label{sec:obs}

\subsection{Observations}
\begin{figure*}
\includegraphics[width=\textwidth]{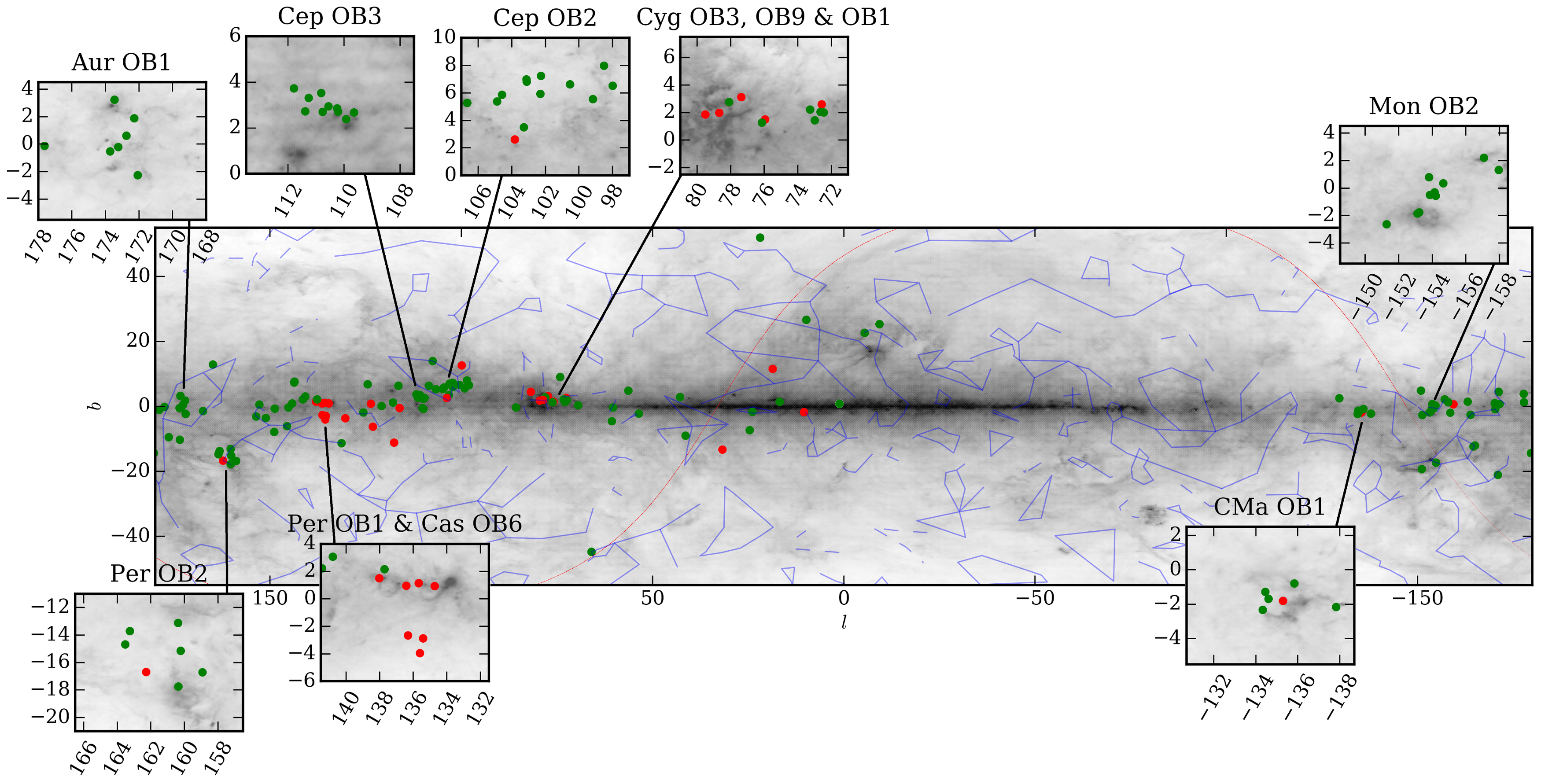}
\caption{Map of the Galaxy with observed stars. Red symbols show LOS where we detect double Na~I and Ca~II profiles with two distinct radial velocity components. Rest of the LOS are marked green. More densely populated OB associations are shown in zoomed-in insets. Red line is the celestial equator. The background picture is the Planck dust map \citep{planck15}.}
\label{fig:map}
\end{figure*}

Spectra were obtained with the 1.82~m telescope of the Observatory of Padua in Asiago, Italy and an echelle-type spectrograph. Observations were done in 17 observing sessions between January 16 2011 and October 15 2016. The spectrograph covers wavelengths between 3700~{\AA} to 7300~{\AA} in 34 echelle orders without any gaps. The resolving power of the setup used is between 20,000 and 23,000. The variation in the resolving power is due to seeing constraints; the slit width was set to 200~$\mu$m in good seeing and to 250~$\mu$m when the seeing was worse. Some of the data was also taken in the shared time with other projects that required the wider slit. Enough signal was collected for each star to have the signal-to-noise ratio (SNR) of 300 or more per pixel over the whole wavelength range where DIBs included in this study are present. A required exposure time to accumulate the required SNR is between 1 minute for brightest stars and 1.5 hours for darkest stars in this study, assuming a typical seeing of $\sim$2 arc seconds. Exposure time was adapted to weather conditions accordingly and the darkest stars were only observed in optimal conditions. A sub-sample of the observed stars has already been analysed in \citet{kos13} with less rigorous methods, so they are all re-analysed here. 5 stars analysed in that paper were re-observed later for reasons unrelated to this paper. In such cases we combined the spectra observed over multiple epochs. Map of the observed stars is shown in Figure \ref{fig:map}.

\subsection{Target selection}
A selection function for program stars was kept simple: a magnitude limit of V=8.5 (darker stars would require infeasible exposure times) and a color cut at spectral type B3 was used. Cooler stars have a spectrum populated by more stellar absorption lines and are unsuitable for the presented study with the approach described further in the text. The color cut was chosen in a way that enough candidates passed the cut. We showed that the color cut is adequate in \citet{kos13}. Further candidates were discarded for not having reliable reddening measurements in the literature. Around 300 candidates remained and we observed 145 of them. The color excess of observed stars varies between just measurable level of few hundreds of magnitude to around E(B-V)=1.2 magnitude. Stars with higher color excess would be too faint at the blue end of the wavelength range and would not fit the accessible magnitude range. We tried to keep the distribution of the color excess of the observed stars as uniform as possible in order to sample more diverse regions of the Galaxy.

\begin{figure}
\includegraphics[width=0.95\columnwidth]{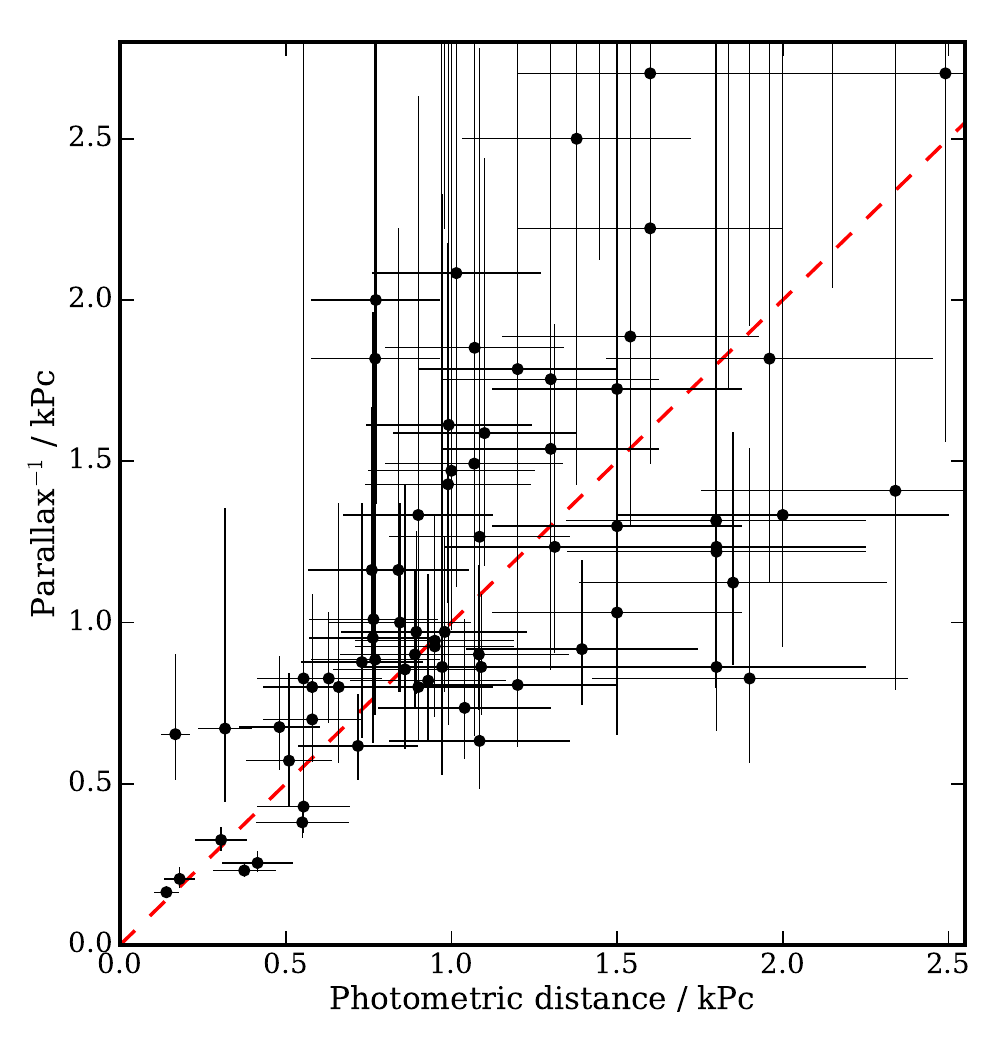}
\caption{Comparison of photometric and trigonometric distances for stars with both distances available. Four stars have trigonometric distances larger than 3~kpc or negative parallax and are missing from the plot. Their errorbars are still shown if they reach into the plotted range. 1:1 relation is shown with a dashed line.}
\label{fig:photrig}
\end{figure}

Four of the observed stars are colder than type B3 stars. They are all rapid rotators \citep{rot05}, so their spectrum is appropriate for our analysis because the rotationally broadened lines satisfy the same criteria as temperature broadened lines of hotter stars.

Distances of targeted stars are taken from \citet{neckel80} (photometric distances) and \citet{gaia} (trigonometric distances). Photometric distance is available for almost every star and the trigonometric distance is available for only 60\% of the stars. Figure \ref{fig:photrig} shows the comparison between the two distances. For most stars both distances match within the errorbars. There are however a few stars where two distances are several sigmas apart. In the analysis we will use the weighted averages between the photometric and trigonometric distances wherever it is possible. The exception are the stars that have a negative mean parallax.

Table \ref{tab:list} lists all relevant information about the observed stars.

\subsection{Reduction}

Spectra were reduced by standard IRAF routines within \texttt{echelle} and \texttt{noao} packages. Images were bias corrected and corrected for cosmic rays. We remove the cosmic rays from the raw images, because most often we only recorded two spectra, so the cosmic rays can not be removed by a median, for example. Spectral traces were then optimally extracted from flat field image and were flat field corrected. Scattered light has been removed in the process. Wavelength calibration has been done against spectra of a Thorium-Argon arc lamp taken with every set of science images. Echelle spectra were normalized and echelle orders were stitched together into a single spectrum. Two or more exposures were made for each star, depending on the required SNR and weather conditions. Single spectra were shifted into the barycentric velocity frame and combined into one spectrum per star. 

An error spectrum tracing the uncertainties for each pixel was created along with the reduced spectrum.

\renewcommand{\arraystretch}{0.96}
\begin{table*}
\begin{tabular}{l c c x{1.3cm} c c x{1.62cm} x{1.4cm} x{1.35cm} x{1.3cm}}
\hline\hline
Name     & $l$      & $b$      & Spectral Type & $V$ & $E(B-V)$ & Photometric Distance$^\mathrm{n}$ & Parallax$^\mathrm{g}$& Observing session & Peak S/N\\\hline
         & $(^\circ)$ & $(^\circ)$ &  & (mag) & (mag) & (kpc) & (mas) & & \\\hline
HD108    & 117.9268 & 1.2498   & O4          & 7.40 & 0.49$^\mathrm{a}$   & 1.8 & 0.82$\pm$0.34  & 14     & 490\\
HD1337   & 117.5880 & -11.0884 & O9II        & 6.14 & 0.34$^\mathrm{a}$   & 1.2 & ?  & 11     & 500\\
HD2905   & 120.8361 & 0.1351   & B1I         & 4.16 & 0.33$^\mathrm{a, v}$& 0.951 & ? & 9      & 820\\
HD5005   & 123.1232 & -6.2436  & O5.5        & 7.76 & 0.41$^\mathrm{s}$   & 2.262 & ? &2      & 520\\
HD5551   & 123.7079 & 0.8524   & B1.5I       & 7.76 & 0.8$^\mathrm{a}$    & 1.5   & 0.58$\pm$0.27 &15     & 520\\
HD6675   & 124.4848 & 6.8684   & B0.5III     & 6.94 & 0.59$^\mathrm{a}$   & 0.9   & 0.75$\pm$0.37 &9      & 550\\
HD7252   & 125.6820 & -1.8678  & B1V         & 7.19 & 0.35$^\mathrm{a}$   & 0.84  & 0.86$\pm$0.41 &15     & 650\\
HD10516  & 131.3247 & -11.3301 & B2V         & 4.09 & 0.264$^\mathrm{m}$  & 0.320 & ? &1      & 540\\
HD10898  & 130.3637 & -3.5869  & B2I         & 7.40 & 0.52$^\mathrm{a}$   & 1.9   & 0.24$\pm$0.28 & 11     & 600\\
HD14818  & 135.6167 & -3.9333  & B2I         & 6.26 & 0.48$^\mathrm{a}$   & 2.0   & 0.30$\pm$0.41& 9, 15  & 520\\
HD14956  & 135.4234 & -2.8622  & B2I         & 7.24 & 0.89$^\mathrm{n, v, a}$ & 1.837 & 0.32$\pm$0.26& 6  & 520\\
HD15558  & 134.7243 & 0.9248   & O4.5III     & 7.87 & 0.84$^\mathrm{v}$   & 1.799 & 0.76$\pm$0.49& 15     & 500\\
HD15690  & 136.3205 & -2.6557  & B2I         & 8.02 & 0.83$^\mathrm{a}$   & 1.6   & 0.45$\pm$0.22 & 14     & 470\\
HD16310  & 136.4230 & 0.9483   & B1II        & 8.10 & 0.92$^\mathrm{a}$   & 1.1   & 0.63$\pm$0.22& 9      & 450\\
HD16429  & 135.6779 & 1.1453   & O9.5I       & 7.85 & 0.92$^\mathrm{a}$   & 1.094 & ? & 6, 14  & 530\\
HD18326  & 138.0260 & 1.5002   & O6.5V       & 7.94 & 0.69$^\mathrm{a}$   & 1.6   & 0.37$\pm$0.23& 11     & 490\\
HD18352  & 137.7273 & 2.1610   & B1V         & 6.85 & 0.47$^\mathrm{a, v}$& 0.58  & 1.25$\pm$0.33& 11     & 650\\
HD20365  & 145.6002 & -6.0578  & B3V         & 5.16 & 0.14$^\mathrm{a}$   & 0.19  & ? & 9      & 970\\
HD20898  & 140.8016 & 3.0642   & B2III       & 8.00 & 0.68$^\mathrm{a}$   & 0.77  & 0.55$\pm$0.33& 14     & 530\\
HD20959  & 141.4407 & 2.2263   & B3III       & 8.05 & 0.47$^\mathrm{a}$   & 0.77  & 1.13$\pm$0.27& 15     & 540\\
HD21455  & 148.9316 & -7.8020  & B7V         & 6.23 & 0.26$^\mathrm{a}$   & 0.14  & 6.11$\pm$0.55 & 14     & 730\\
HD22253  & 144.2776 & 0.9244   & B0.5III     & 6.55 & 0.61$^\mathrm{a}$   & 0.718 & 1.62$\pm$0.33& 3 & 560\\
HD22298  & 145.2364 & -0.3176  & B2V         & 7.69 & 0.68$^\mathrm{s}$   & ?     & 1.15$\pm$0.27& 11     & 480\\
HD22951  & 158.9198 & -16.7030 & B0.5V       & 4.98 & 0.263$^\mathrm{m}$  & 0.402 & ? & 1 & 420\\
HD23180  & 160.3637 & -17.7398 & B1III       & 3.85 & 0.29$^\mathrm{v,l}$ & 0.289 & ? & 1, 3 & 520\\
HD24131  & 160.2266 & -15.1369 & B1V         & 5.78 & 0.27$^\mathrm{m}$   & 0.554 & 2.33$\pm$0.53 & 1 & 390\\
HD24398  & 162.2892 & -16.6904 & B1I         & 2.88 & 0.34$^\mathrm{l}$   & 0.352 & ? & 1, 3 & 560\\
HD24431  & 148.8384 & -0.7085  & O9III       & 6.74 & 0.69$^\mathrm{a}$   & 0.95  & 1.08$\pm$0.33& 9      & 390\\
HD24432  & 151.1206 & -3.4953  & B3II        & 6.87 & 0.74$^\mathrm{a, v}$& 1.278 & ? & 10     & 600\\
HD24912  & 160.3723 & -13.1065 & O7.5III     & 4.06 & 0.32$^\mathrm{a, v}$& 0.421 & ? & 9      & 990\\
HD25443  & 143.6820 & 7.3544   & B0.5III     & 6.78 & 0.57$^\mathrm{a, v}$& 1.378 & 0.40$\pm$0.30 & 11     & 550\\
HD25539  & 163.5201 & -14.6703 & B3V         & 6.87 & 0.287$^\mathrm{m}$ & 0.415 & 3.93$\pm$0.46& 1 & 420\\
HD25638  & 143.6705 & 7.6576   & B0III       & 6.93 & 0.73$^\mathrm{v,s}$ & 1.085 & 1.58$\pm$0.48 & 3& 610\\
HD25639  & 143.6755 & 7.6583   & B0III       & 6.86 & 0.73$^\mathrm{s, a}$ & 1.085 & 0.79$\pm$0.43& 3 & 520\\
HD25833  & 163.2454 & -13.7164 & B3V         & 6.69 & 0.235$^\mathrm{m}$ & ? & ? & 1 & 430\\
HD25940  & 153.6543 & -3.0450  & B3V         & 4.00 & 0.173$^\mathrm{s}$ & 0.096 & ? & 6 & 540\\
HD27192  & 152.8033 & 0.5738   & B1.5IV      & 5.55 & 0.24$^\mathrm{a}$   & 0.5  & ? & 9      & 740\\
HD30614  & 180.3423 & -14.3712 & O9.5I       & 4.30 & 0.315$^\mathrm{a, s, v}$ & 0.813 & ? & 6 & 540\\
HD30675  & 173.6059 & -10.2050 & B3V         & 7.55 & 0.525$^\mathrm{v, a}$& 0.375 & 4.33$\pm$0.38& 14     & 550\\
HD30870  & 189.0327 & -20.9981 & B5V         & 6.11 & 0.24$^\mathrm{a}$   & 0.18 &4.88$\pm$0.71& 14     & 620\\
HD32018  & 176.4740 & -9.4588  & B2IV        & 7.52 & 0.80$^\mathrm{n, s}$ & 0.429 &? & 6 & 520\\
HD32672  & 167.5261 & -1.3526  & B2IV        & 7.77 & 0.41$^\mathrm{n, s}$ & 0.9 &1.25$\pm$0.32& 14, 15 & 280\\
HD34078  & 172.0813 & -2.2592  & O9.5V       & 5.96 & 0.52$^\mathrm{v, l, a}$ & 0.55 &2.63$\pm$0.36& 13     & 540\\
HD35633  & 173.2372 & -0.1957  & B0.5IV      & 8.09 & 0.6$^\mathrm{a}$ & 1.3 &0.57$\pm$0.28& 15     & 460\\
HD35653  & 173.7297 & -0.5141  & B0.5V       & 7.50 & 0.4$^\mathrm{a}$ & 1.0 &0.68$\pm$0.34& 12     & 390\\
HD35921  & 172.7615 & 0.6106   & O9.5III     & 6.85 & 0.63$^\mathrm{n}$ & 1.562 &? & 3 & 490\\
HD36483  & 172.2940 & 1.8776   & O9.5IV      & 8.24 & 0.73$^\mathrm{a}$ & 1.7 &? & 14     & 500\\
HD36822  & 195.4023 & -12.2891 & B0III       & 4.41 & 0.14$^\mathrm{a}$ & 0.56 & ? & 11     & 820\\
HD36861  & 195.0520 & -11.9949 & O8III       & 3.30 & 0.120$^\mathrm{s}$ & 0.372 & ?& 6 & 540\\
HD37061  & 208.9248 & -19.2736 & O9V         & 6.83 & 0.54$^\mathrm{v, a}$ & 0.46 & ? & 14     & 530\\
HD37128  & 205.2121 & -17.2416 & B0I         & 1.70 & 0.081$^\mathrm{s}$ & 0.356 & ? & 6 & 550\\
HD37366  & 177.6302 & -0.1136  & O9.5IV      & 7.64 & 0.39$^\mathrm{a}$ & 1.5 & 0.97$\pm$0.56& 14     & 430\\
HD37367  & 179.0360 & -1.0258  & B2IV        & 5.99 & 0.41$^\mathrm{v, a}$ & 0.32 &? & 12     & 530\\
HD37737  & 173.4643 & 3.2435   & O9.5II      & 8.06 & 0.61$^\mathrm{a}$ & 2.0 & 0.75$\pm$0.33& 11     & 480\\
HD39746  & 182.1813 & 1.2746   & B1II        & 7.04 & 0.48$^\mathrm{a}$ & 1.2 &0.56$\pm$0.32& 13     & 450\\
HD41117  & 189.6918 & -0.8604  & B2I         & 4.63 & 0.425$^\mathrm{a,v}$ & 0.95 & ? & 7      & 530\\
HD41161  & 164.9727 & 12.8935  & O8V         & 6.76 & 0.21$^\mathrm{a}$ & 1.8 & ? & 14     & 590\\
HD41398  & 182.2568 & 3.8739   & B2I         & 7.49 & 0.49$^\mathrm{a}$ & 2.1 & ? & 12     & 410\\
HD41690  & 188.6036 & 0.7466   & B1V         & 7.71 & 0.47$^\mathrm{a,v}$ & 0.87 &? & 13     & 430\\
\hline
\end{tabular}
\caption{Basic properties of observed stars. Observing sessions: 1 (1. 16. 2011), 2 (9. 15. 2011 -- 9. 16. 2011), 3 (2. 3. 2012 -- 2. 6. 2012), 4 (5. 3. 2012 -- 5. 5.2012), 5 (8. 29. 2012 -- 9. 5. 2012), 6 (24. 11. 2012 -- 26. 11. 2012), 7 (24. 4. 2013 -- 25. 4. 2013), 8 (24. 6. 2013 -- 26. 6. 2013), 9 (21. 9. 2013 -- 23. 9. 2013), 10 (18, 10, 2013 -- 18. 10. 2013), 11 (12. 1. 2014 -- 13. 1. 2014), 12 (19. 3. 2014 -- 21. 3. 2014), 13 (10. 4. 2014 -- 13. 4. 2014), 14 (3. 1. 2015 -- 5. 1. 2015), 15 (31. 1. 2015 -- 3. 2. 2015), 16 (1. 4. 2015 -- 3. 4. 2015), 17 (15. 10. 2016). References: v: \citet{valencic04}, s: \citet{snow77}, a: \citet{savage85}, m: \citet{maeder75}, l: \citet{slyk06}, n: \citet{neckel80}, f: \citet{fitzpatrick07}, g: \citet{gaia}. Coordinates, magnitudes and spectral types are taken from Simbad. Uncertainties of photometric distances are not given explicitly for each star but are estimated to be between 16 and 25\%\citep{neckel80}. Typical uncertainty for $\mathrm{E(B-V)}$ is 0.03~mag. Peak S/N gives the signal-to-noise ratio in the region with highest photon counts, usually around 5800~{\AA}}
\label{tab:list}
\end{table*}

\begin{table*}
\addtocounter{table}{-1}
\begin{tabular}{l c c x{1.3cm} c c x{1.62cm} x{1.4cm} x{1.35cm} x{1.3cm}}
\hline\hline
Name     & $l$      & $b$      & Spectral Type & $V$ & $E(B-V)$ & Photometric Distance$^\mathrm{n}$ & Parallax$^\mathrm{g}$& Observing session & Peak S/N\\\hline
         & $(^\circ)$ & $(^\circ)$ &  & (mag) & (mag) & (kpc) & (mas) & & \\\hline
HD42352  & 196.1555 & -2.5736  & B1III       & 6.93 & 0.26$^\mathrm{a}$ & 1.2 & 1.24$\pm$0.38& 14     & 620\\
HD43818  & 188.8364 & 4.4803   & B0II        & 6.92 & 0.59$^\mathrm{v, a}$ & 1.312 & 0.81$\pm$0.29& 6, 12 & 510\\
HD45314  & 196.9593 & 1.5246   & O9          & 6.64 & 0.44$^\mathrm{a, v, l}$ & 0.86 & 1.17$\pm$0.47& 14     & 620\\
HD45995  & 200.6231 & 0.6784   & B2V         & 6.14 & 0.168$^\mathrm{m}$ & 0.410 & ? & 1 & 380\\
HD46150  & 206.3058 & -2.0695  & O5V         & 6.73 & 0.45$^\mathrm{v, s, a, n}$ & 1.4 & 1.06$\pm$0.39 & 17 & 330\\
HD46202 & 206.3134 & -2.0035   & O9.2V       & 8.19 & 0.48$^\mathrm{v, s, a, n}$ & 1.9 & 0.76$\pm$0.31 & 17 & 280\\
HD46484  & 206.7824 & -1.7606  & B2I         & 7.74 & 0.62$^\mathrm{a}$ & 0.76 & 0.86$\pm$0.26& 15     & 550\\
HD46485  & 206.8975 & -1.8363  & O7V         & 8.27 & 0.63$^\mathrm{a, v}$ & 1.85 & 0.89$\pm$0.26& 15     & 400\\
HD46573  & 208.7297 & -2.6311  & O7V         & 7.93 & 0.66$^\mathrm{a}$ & 1.8 &0.81$\pm$0.28& 11     & 490\\
HD46883  & 202.0472 & 1.3230   & B0.5V       & 7.78 & 0.63$^\mathrm{a,v}$ & 0.845 & 1.00$\pm$0.27& 14     & 480\\
HD46966  & 205.8096 & -0.5491  & O8.5IV      & 6.87 & 0.27$^\mathrm{a,v}$ & 1.578 & ? & 12     & 460\\
HD47107  & 206.1481 & -0.4979  & B1V         & 8.00 & 0.23$^\mathrm{f}$ & ? & 0.01$\pm$0.71& 16     & 480\\
HD47129  & 205.8740 & -0.3111  & O8I         & 6.06 & 0.37$^\mathrm{a,v}$ & 0.945 & ? & 12     & 500\\
HD47417  & 205.3523 & 0.3492   & B0IV        & 6.95 & 0.30$^\mathrm{a, v}$ & 1.22 & ? & 16     & 570\\
HD47839  & 202.9363 & 2.1985   & O7V         & 4.64 & 0.07$^\mathrm{a}$ & 0.93 & ? & 14     & 940\\
HD48099  & 206.2096 & 0.7982   & O5V         & 6.37 & 0.265$^\mathrm{a, v}$ & 1.56 & ? & 16     & 600\\
HD52382  & 222.1707 & -2.1549  & B0.5I       & 6.57 & 0.4$^\mathrm{a}$ & 1.5 & 0.77$\pm$0.36& 11     & 540\\
HD52559  & 209.1666 & 4.8565   & B2IV        & 6.58 & 0.22$^\mathrm{a}$ & 0.69 & ? & 14     & 650\\
HD53974  & 224.7096 & -1.7938  & B2I         & 5.39 & 0.32$^\mathrm{a, v}$ & 0.50 & ? & 12     & 550\\
HD53975  & 225.6786 & -2.3157  & O7.5V       & 6.50 & 0.21$^\mathrm{a}$ & 1.6 & ? & 12     & 460\\
HD54439  & 225.3979 & -1.6790  & B2II        & 7.68 & 0.285$^\mathrm{a,v}$ & 1.9 & 1.21$\pm$0.56& 16     & 370\\
HD54662  & 224.1685 & -0.7784  & O7V         & 6.21 & 0.35$^\mathrm{a}$ & 1.3 &0.65$\pm$0.52& 12     & 490\\
HD54879  & 225.5511 & -1.2762  & O9.7V       & 7.65 & 0.3$^\mathrm{a}$ & 1.8 &1.16$\pm$0.34& 15     & 560\\
HD60325  & 230.4540 & 2.5196   & B2II        & 6.21 & 0.22$^\mathrm{a}$ & 0.66 &1.25$\pm$0.52& 11     & 570\\
HD137569 & 21.8661  & 51.9303  & B5III       & 7.91 & 0.40$^\mathrm{a}$ & 0.564 &-0.37$\pm$0.44& 4 & 460\\
HD142096 & 350.7243 & 25.3801  & B3V         & 5.03 & 0.205$^\mathrm{a, n}$ & 0.18 &?& 8      & 210\\
HD145502 & 354.6087 & 22.7002  & B2V         & 4.00 & 0.276$^\mathrm{a, n}$ & 0.12 &? & 16     & 510\\
HD149363 & 9.8524   & 26.6906  & B1I         & 7.81 & 0.308$^\mathrm{a, n}$ & 1.96 & 0.55$\pm$0.34& 12     & 400\\
HD161056 & 18.6702  & 11.5808  & B1.5V       & 6.30 & 0.63$^\mathrm{a}$ & 0.167 & 1.53$\pm$0.42& 3 & 550\\
HD167264 & 10.4557  & -1.7408  & O9.7I       & 5.37 & 0.30$^\mathrm{a, v, n}$ & 1.40 & ? & 13     & 660\\
HD167451 & 16.8262  & 1.5203   & B1I         & 8.30 & 1.01$^\mathrm{n}$ & 1.540 & 0.53$\pm$0.24& 2 & 480\\
HD172488 & 23.9639  & -1.6220  & B0I         & 7.62 & 0.825$^\mathrm{a, n}$ & 0.63 &1.21$\pm$0.24& 7      & 310\\
HD176304 & 42.8531  & 2.8820   & B2V         & 6.75 & 0.49$^\mathrm{a}$ & 0.317 & 1.49$\pm$0.75& 4, 13 & 490\\
HD176853 & 24.5945  & -7.3127  & B2V         & 6.64 & 0.46$^\mathrm{n}$ & 0.305 & 3.07$\pm$0.32& 2 & 470\\
HD180968 & 56.3582  & 4.8532   & B0.5IV      & 5.48 & 0.27$^\mathrm{a, n, v}$ & 0.54 &? & 12     & 570\\
HD184915 & 31.7709  & -13.2866 & B0.5III     & 4.96 & 0.275$^\mathrm{a, n}$ & 0.61 & ? & 13     & 700\\
HD185418 & 53.6024  & -2.1709  & B0.5V       & 7.52 & 0.50$^\mathrm{a, n, v}$ & 0.93 & 1.22$\pm$0.35& 16     & 510\\
HD185423 & 41.4269  & -8.9437  & B5I         & 6.36 & 0.245$^\mathrm{a, n}$ & 0.51 & ? & 16     & 620\\
HD185780 & 74.1711  & 9.0540   & B0III       & 7.72 & 0.235$^\mathrm{a, n}$ & 2.51 & ? & 13     & 550\\
HD186841 & 60.4065  & -0.2918  & B1I         & 8.01 & 0.97$^\mathrm{a, n}$ & 2.02 & 0.10$\pm$0.23& 13     & 400\\
HD190066 & 60.6853  & -4.5385  & B1I         & 6.60 & 0.37$^\mathrm{a, n}$ & 2.34 & 0.71$\pm$0.55& 13     & 539\\
HD190429 & 72.5852  & 2.6136   & O4I         & 6.63 & 0.48$^\mathrm{a, n}$ & 1.33 & ? & 8      & 510\\
HD190603 & 69.4863  & 0.3895   & B1.5I       & 5.65 & 0.71$^\mathrm{a, n, v}$ & 0.97 & 0.21$\pm$0.72& 8      & 380\\
HD190864 & 72.4673  & 2.0183   & O6.5III     & 7.78 & 0.50$^\mathrm{a, n}$ & 2.15 & 0.26$\pm$0.23& 7      & 610\\
HD190918 & 72.6516  & 2.0651   & W-R         & 6.81 & 0.447$^\mathrm{s}$ & 1.015 & 0.48$\pm$0.42& 2 & 540\\
HD191139 & 73.2740  & 2.2208   & B0.5III     & 8.02 & 0.50$^\mathrm{a, n}$ & 1.81 & -0.19$\pm$0.27& 16     & 460\\
HD191612 & 72.9917  & 1.4344   & O8          & 7.80 & 0.58$^\mathrm{a, v, n}$ & 1.45 & -0.02$\pm$0.26& 7      & 520\\
HD192660 & 77.3737  & 3.1375   & B0I         & 7.56 & 0.90$^\mathrm{v,a}$ & 1.447 & 0.25$\pm$0.22& 2, 12 & 510\\
HD193183 & 75.9500  & 1.4992   & B1.5I       & 7.02 & 0.625$^\mathrm{s}$ & 2.491 & 0.37$\pm$0.27& 2 & 530\\
HD193322 & 78.0986  & 2.7807   & O9IV        & 5.84 & 0.415$^\mathrm{a, n, v}$ & 0.72 & ? & 11     & 580\\
HD193443 & 76.1497  & 1.2833   & O9III       & 7.24 & 0.71$^\mathrm{a}$ & 1.238 & ? & 5 & 520\\
HD194057 & 81.8466  & 4.5413   & B1I         & 7.52 & 1.055$^\mathrm{a, n}$ & 0.98 & 0.22$\pm$0.23& 8      & 330\\
HD194279 & 78.6780  & 1.9863   & B1.5I       & 7.09 & 1.21$^\mathrm{a}$ & 0.970 & ? & 5 & 510\\
HD194839 & 79.5171  & 1.8727   & B0.5I       & 7.53 & 1.217$^\mathrm{a}$ & 1.000 & -0.14$\pm$0.27& 2 & 490\\
HD198781 & 99.9447  & 12.6137  & B0.5V       & 6.46 & 0.35$^\mathrm{a, v, n}$ & 0.73 & 1.14$\pm$0.41& 13     & 610\\
HD199579 & 85.6967  & -0.2996  & O6.5V       & 5.96 & 0.37$^\mathrm{a, n, v}$ & 1.07 & 0.54$\pm$0.53& 13     & 600\\
HD202214 & 98.5202  & 7.9852   & O9.5IV      & 5.72 & 0.41$^\mathrm{a, n}$ & 0.61 &? & 14     & 630\\
HD203025 & 97.9974  & 6.5304   & B2III       & 6.43 & 0.453$^\mathrm{s}$ & 0.481 & 1.48$\pm$0.36& 2 & 500\\
HD204827 & 99.1667  & 5.5525   & O9.5IV      & 7.94 & 1.09$^\mathrm{a, n, v, l}$ & 0.58 & 1.43$\pm$0.33& 15     & 440\\
HD205021 & 107.5392 & 14.0261  & B0.5III     & 3.23 & 0.025$^\mathrm{a, s}$ & 0.178 & ? & 12     & 700\\
HD205139 & 100.5455 & 6.6217   & B1II        & 5.54 & 0.39$^\mathrm{s}$ & 0.763 & 1.05$\pm$0.54& 2 & 540\\
HD206165 & 102.2713 & 7.2469   & B2I         & 4.79 & 0.473$^\mathrm{a}$ & 0.576 & ? & 2 & 510\\
HD207198 & 103.1363 & 6.9949   & O8.5II      & 5.94 & 0.59$^\mathrm{a, n, v, l}$ & 1.06 & ? & 16     & 510\\
HD207260 & 102.3099 & 5.9342   & A2I         & 4.31 & 0.53$^\mathrm{n}$ & 1.069 & 0.67$\pm$0.87& 4 & 500\\
HD207308 & 103.1089 & 6.8176   & B0.5V       & 7.49 & 0.53$^\mathrm{s}$ & 0.765 & 0.99$\pm$0.24& 4 & 500\\
HD209339 & 104.5775 & 5.8693   & O9.7IV      & 6.69 & 0.36$^\mathrm{a, n, v}$ & 1.04 & 1.36$\pm$0.37& 16     & 560\\
HD209744 & 103.2795 & 3.4975   & B1V         & 6.71 & 0.48$^\mathrm{s}$ & 0.554 & 1.21$\pm$0.92& 2 & 510\\
\hline
\end{tabular}
\caption{Continued.}
\end{table*}

\begin{table*}
\addtocounter{table}{-1}
\begin{tabular}{l c c x{1.3cm} c c x{1.62cm} x{1.4cm} x{1.35cm} x{1.3cm}}
\hline\hline
Name     & $l$      & $b$      & Spectral Type & $V$ & $E(B-V)$ & Photometric Distance$^\mathrm{n}$ & Parallax$^\mathrm{g}$& Observing session & Peak S/N\\\hline
         & $(^\circ)$ & $(^\circ)$ &  & (mag) & (mag) & (kpc) & (mas) & & \\\hline
HD209975 & 104.8707 & 5.3906   & O9.5I       & 5.11 & 0.35$^\mathrm{a, v}$ & 0.992 &0.62$\pm$0.84& 6 & 550\\
HD210839 & 103.8281 & 2.6107   & O6I         & 5.08 & 0.57$^\mathrm{a}$ & 0.550 & ? & 5 &530\\
HD211880 & 106.6677 & 5.2890   & B0.5V       & 7.75 & 0.605$^\mathrm{a, n}$ & 0.95 & 1.06$\pm$0.25& 10     & 560\\
HD212571 & 66.0068  & -44.7395 & B1V         & 4.79 & 0.233$^\mathrm{s}$ & 0.319 &?& 2 & 530\\
HD213087 & 108.4991 & 6.3878   & B0.5I       & 5.53 & 0.59$^\mathrm{a, n}$ & 0.76 & ? & 14     & 540\\
HD216532 & 109.6520 & 2.6757   & O8.5V       & 8.00 & 0.86$^\mathrm{a, n}$ & 1.09 & 1.16$\pm$0.24& 9      & 500\\
HD216898 & 109.9274 & 2.3930   & O8.5V       & 8.04 & 0.85$^\mathrm{a}$ & 1.394 &1.09$\pm$0.25& 5 & 500\\
HD217035 & 110.2532 & 2.8609   & B0V         & 7.74 & 0.76$^\mathrm{a,s,n}$ & 0.880 & ? & 5 & 510\\
HD217086 & 110.2206 & 2.7197   & O7V         & 7.71 & 0.95$^\mathrm{v,a,s,n}$ & 0.980 & 1.03$\pm$0.24& 4 & 510\\
HD217297 & 110.8177 & 3.5221   & B1.5V       & 7.41 & 0.57$^\mathrm{s,a}$ & 0.774 & ? & 5 & 540\\
HD217312 & 110.5631 & 2.9460   & B0IV        & 7.44 & 0.68$^\mathrm{a}$ & 0.972 & 1.16$\pm$0.73& 4 & 520\\
HD217657 & 110.7685 & 2.7033   & B0.5V       & 8.12 & 0.77$^\mathrm{a}$ & 0.89 & 1.11$\pm$0.25& 14     & 475\\
HD217919 & 111.2657 & 3.3094   & B0IV        & 8.24 & 0.93$^\mathrm{a}$ & 0.894 &1.03$\pm$0.25& 5 & 500\\
HD218323 & 111.7993 & 3.7279   & B0III       & 7.65 & 0.90$^\mathrm{a}$ & 1.083 &1.11$\pm$0.26& 5 & 510\\
HD218342 & 111.3912 & 2.7249   & B5          & 7.43 & 0.715$^\mathrm{a, n}$ & 0.99 &0.70$\pm$0.24& 14     & 490\\
HD218376 & 109.9477 & -0.7834  & B0.5IV      & 4.84 & 0.25$^\mathrm{v,a}$ & 0.382 &? & 5 & 520\\
HD218440 & 110.1336 & -0.5282  & B2V         & 6.42 & 0.215$^\mathrm{a, n}$ & 0.51 & 1.75$\pm$0.56& 15     & 510\\
HD222568 & 116.5030 & 6.3694   & B1IV        & 7.71 & 0.65$^\mathrm{n}$ & 0.772 &0.50$\pm$0.23& 2 & 510\\
HD223987 & 116.1841 & -0.5138  & B0          & 7.60 & 0.705$^\mathrm{a, n}$ & 1.56 &? & 14     & 500\\
\hline
\end{tabular}
\caption{Continued}
\end{table*}

\section{Methods}
\label{sec:method}

\subsection{DIB profiles}

Different DIBs have a diverse range of profiles: a few are symmetrical single lines with no resolved substructure \citep{sarre95}, most are bands with a resolved substructure \citep{gala02}, and sometimes they are blended individual DIBs \citep{jenniskens94}. Data used in this work have a resolving power just above R=20,000, which is not enough to resolve any substructure. The substructure or blended DIBs, however, can distort the profiles into asymmetric shapes. This is detectable in our spectra. We therefore describe the DIB profile with a Gaussian:
\begin{equation}
\Phi(\lambda;A,\lambda_0,\sigma)=A\frac{1}{\sqrt{2\pi}\sigma}\exp \left( \frac{-(\lambda-\lambda_0)^2}{2\sigma^2}\right)
\end{equation}
where the possible asymmetry ($asym$) of the profile is introduced by making the width $\sigma$ a function of the wavelength:
\begin{equation}
\sigma(\lambda;\lambda_0,asym)=\frac{2\sigma}{1+\exp\left( asym (\lambda-\lambda_0)\right)}
\end{equation}
Such implementation of the asymmetry is just one of the possibilities and since there is no theoretical prescription for the asymmetry, our choice is the one that works best in the fitting scheme. See \citet{kos13} for a comparison of two more asymmetry implementations. 

All DIBs studied here are optically thin, so their equivalent widths (EWs) are linearly proportional to the column densities of the absorbing material. Column densities, however, cannot be calculated, because the oscillator strengths of DIBs are unknown. The opposite holds for atomic and molecular absorption lines studied in this work. They can be optically thick, so the column density must be used to describe the amount of the ISM in the LOS. DIB strengths are therefore expressed as equivalent widths and atomic and molecular abundances as column densities. The same schema is used for fitting DIBs, molecular, and atomic absorption profiles. We will describe the absorption lines with an absorption line shape $I(\lambda)$:
\begin{equation}
I(\lambda)=\exp(-\tau(\lambda)),
\end{equation}
where
\begin{equation}
\tau(\lambda)=N\frac{\pi e^2}{m_ec}f_{lu}\Phi(\lambda),
\end{equation}
where $N$ is the column density of the absorber, and $f_{lu}$ is the oscillator strength of the observed transition. The strength of a spectral line is measured by the equivalent width (assuming a normalized spectrum):
\begin{equation}
EW=\int_0^\infty \left(1-I\left(\lambda\right)\right)\mathrm{d}\lambda.
\end{equation}

\subsection{Multi-component profiles}

Some of the observed spectra show interstellar absorption lines with two components with different radial velocities. This means that the LOS passes two distinct clouds of the ISM. We detected 30 such LOS. It is possible that some of the undetected ones also pass more than one distinct cloud of the ISM but remain undetected, if the clouds have the radial velocity with a difference lower than our resolution. In no cases we found more than two components.

To detect the multi-component LOS we use Na~I and Ca~II lines (there are two of each in our wavelength range) which are detectable in every spectrum. Atomic lines, unlike DIBs, are unresolved, so they are as narrow as the resolution allows. It is possible that Na~I and Ca~II doublets are naturally broadened by cloud rotation or turbulence, but it is highly unlikely that any of these velocities would reach 15~$\mathrm{km\ s^{-1}}$, which is approximately our resolution.

We fitted the Na~I and Ca~II doublets with a profile that allowed the second component within a plausible velocity range and strength ratio. Na and Ca profiles were fitted separately, but both lines in each doublet were fitted at the same time. All LOS with strength ratios larger than 0.1 and line separation larger than 15~$\mathrm{km\ s^{-1}}$ were marked possible multi-component. If the fitted values for the Ca doublet and the Na doublet agreed to 10~$\mathrm{km\ s^{-1}}$ or better in line separation and 50\% or better for the strength ratio, the LOS was considered multi-component. A mild criterion for the strength ratio is due to the fact that different clouds can have different abundances of each element.

As it is evident from \citet{kos13} or later in this paper, some DIBs have a very poor correlation between their EWs and strengths of Ca and Na lines. Therefore we can not use the double profiles we calculate from Na and Ca lines as templates for DIBs. Some DIBs like DIB 5780\footnote{We name DIBs by their wavelengths rounded to the nearest {\AA}. A DIB at 5780~{\AA} is therefore called DIB 5780. We use the same name for the molecule that produces the DIB (the carrier), as the actual molecules are usually not known. ''DIB 5780'' might therefore refer to the molecule and not to the absorption feature, depending on the context.} that are the key to the conclusions of this paper can hardly be fitted with multiple components, because they are much broader than the separation between the components. Such fits would be degenerate and would not provide any more information than a single component fit. We will therefore use LOS with multiple components qualitatively only.

\subsection{Fitting line profiles with Gaussian processes}

DIBs are observed in spectra of hot stars with few spectral lines of their own. Spectral lines of hot stars are wide and resolved, but still pose a problem when weak DIBs are observed with amplitudes of an order of 1\% below the continuum in regions populated by stellar lines. Using any of the libraries of synthetic spectra to remove the stellar component from our spectra does not solve the problem, as no model of stellar atmospheres is able to predict the spectrum with sub-one-percent accuracy. 

Accounting for stellar spectral features is trivial in some cases, as they are wider than DIBs and a local continuum around the DIB can be estimated. This approach requires a certain level of human input and the estimates for the continuum are subjective. It can also introduce systematic errors and complicates the estimation of uncertainties. In a case where the stellar spectral features are of the same width as DIBs, such approach can produce ambiguous measurements of DIB profiles. 

In addition to stellar spectral features the signal is affected by a correlated noise originating from the steps we do in the data reduction. The correlation varies with wavelength and it is also not smooth, as the final spectrum is produced from combined and overlapping echelle orders. Such correlated noise brings additional complications to calculation of the uncertainties of measured quantities.

We solve both of the above problems (spectra polluted by stellar spectral features and correlated noise) in one step by using Gaussian processes to fit stellar spectral features and correlated noise at the same time as we fit a DIB profile to the spectra.

Only a 20~{\AA} wide region around each DIB or an ISM absorption line is used to fit the profile, so any spectral features with a scale $>20$~{\AA} are irrelevant.

\subsubsection{Gaussian processes} 

The Gaussian process is a statistical model, where observations occur in a continuous domain. Each data-point is a normal distribution and each collection of data-points is a multivariate normal distribution \citep{gp05}. The Gaussian process can be used as a machine learning algorithm to predict values where no data is observed and also predict the uncertainties. We use Gaussian processes as a regression engine. 

We want to fit a function $f_\Theta(x)$ to $N$ data-points $(x_n,y_n)$ with uncertainties $\sigma_n^2$, where $\Theta$ are free parameters. We can write the ln-likelihood as:
\begin{equation}
\ln p(y_n|x_n,\sigma_n^2,\Theta)=-\frac{1}{2}\sum_{n=1}^N \frac{(y_n-f_\Theta(x_n))^2}{\sigma_n^2}+C,
\end{equation}
which can be rewritten as a matrix equation:
\begin{equation}
\ln p(y_n|x_n,\sigma_n^2,\Theta)=-\frac{1}{2}\mathbf{r}^TK^{-1}\mathbf{r}-\frac{1}{2}\ln \det(K)-\frac{N}{2}\ln 2\pi,
\end{equation}
where $\mathbf{r}$ is the residual vector $\mathbf{r}=(y_n-f_\Theta(x_n))$ of length $N$ and $K$ is the covariance matrix of size $N\times N$:
\begin{equation}
K=\left(
\begin{array}{cccc}
\sigma_1^2 & 0 & \cdots & 0\\
0 & \sigma_2^2 & \cdots & 0\\
\vdots & \vdots & \ddots & \vdots\\
0 & 0 & \cdots & \sigma_N^2
\end{array}
\right)
\end{equation}
The covariance matrix above is diagonal, so taking into the account white noise only. If we want to introduce correlated noise, we simply populate off-diagonal elements with non-zero values. In our case we want to fit the correlated noise, which would be impossible if every element in the covariance matrix was a free parameter. Instead we model the covariance matrix with a covariance function:
\begin{equation}
K_{ij}=\sigma_i^2 \delta_{ij}+k(x_i,x_j),
\end{equation}
where $\delta_{ij}$ is the Kronecker delta and $k(x_i,x_j)$ is a covariance function. In our case the covariance functions will be functions of only two parameters.

We are left with maximizing the ln-likelihood to get the optimal parameters $\Theta$ of the $f_\Theta(x)$ function. The parameters of the covariance function can also be fitted, if the noise characteristics are not known.

\subsubsection{Covariance functions}

We use two different covariance functions (or kernels) to describe two components of the correlated noise. The exponential-squared covariance function models the wide stellar absorption lines:
\begin{equation}
k_{se}(x,x')=a\exp\left(\frac{||x-x'||^2}{2l^2}\right),
\label{eq:k1}
\end{equation}
and a Mat\'{e}rn $3/2$ covariance function models the correlated noise:
\begin{equation}
k_{m3/2}(x,x')=a\left(1+\frac{\sqrt{3}||x-x'||}{l}\right)\exp\left(-\frac{\sqrt{3}||x-x'||}{l}\right),
\label{eq:k2}
\end{equation}
where $l$ is a characteristic width of each covariance function and $a$ is the scaling. Many other covariance functions are commonly used and some are compared in \citet{gp05}.

\begin{figure}
\includegraphics[width=\columnwidth]{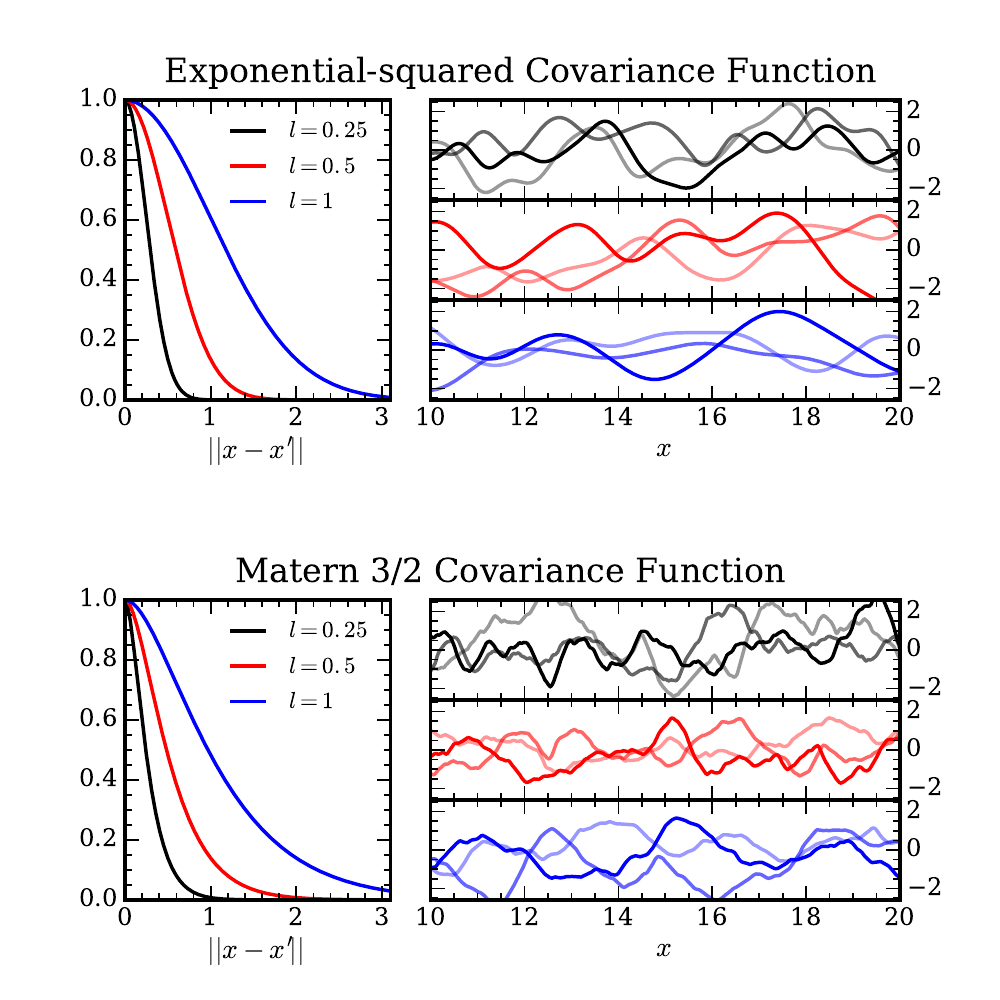}
\caption{Top: Exponential-squared covariance functions with three different values of the characteristic width $l$ are plotted in the left panel. Three random functions drawn from a Gaussian process with the exponential-squared covariance function are plotted on the right for a different $l$ in each of the three panels. Bottom: Mat\'{e}rn $3/2$ covariance functions with three different values of the characteristic width $l$ are plotted in the left panel. Three random functions drawn from a Gaussian process with the Mat\'{e}rn $3/2$ covariance function are plotted on the right for a different $l$ in each of the three panels.}
\label{fig:kernels}
\end{figure}

Figure \ref{fig:kernels} shows exponential-squared and Mat\'{e}rn $3/2$ covariance functions and a few samples of random functions $f \sim \mathcal{GP}(0,k(x,x'))$ constructed from each covariance function. $a$ equals to 1 in the given examples.

Scalings $a$ in front of each covariance function in equations \ref{eq:k1} and \ref{eq:k2} can be estimated immediately and does not have to be fitted. $a$ scales as the variance of the noise we want to describe with the given covariance function. One used to model the stellar component is scaled to the variance of the spectrum in the wavelength range that we are interested in, and the second one, used to model the noise, is scaled to the squared average uncertainty in the wavelength range of interest. The final covariance function entering the maximization of the ln-likelihood is a sum of a Mat\'{e}rn $3/2$ and exponential-squared covariance functions.

Characteristic widths $l$ remain to be fitted, as they vary with the stellar type and noise characteristics. We fit the free parameters with a Bayesian MCMC scheme \citep{fm13}.

\begin{figure}
\includegraphics[width=\columnwidth]{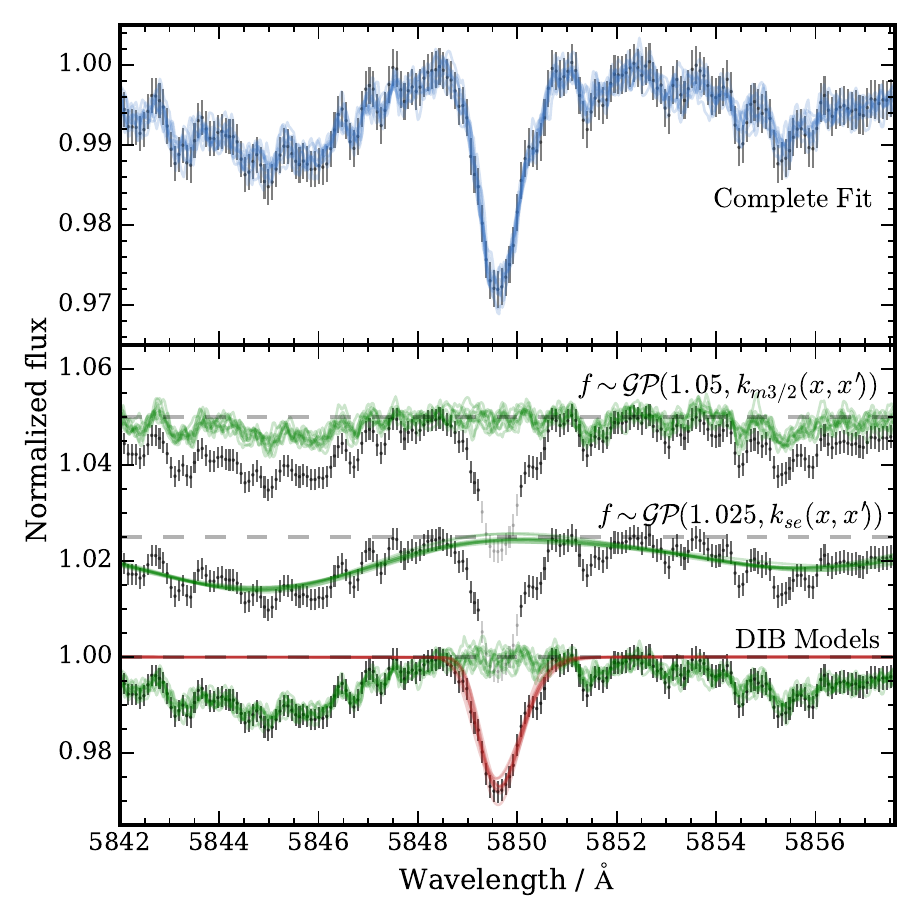}
\caption{DIB 5850 in a spectrum of star HD 18326 with a fitted DIB and noise profile. Top: The complete fit. Black points show the measured flux and measured uncertainties. In blue are ten random samples from the posterior distribution. Bottom: Decomposition of the blue curves from the top panel.  In green are random functions sampled from the Gaussian process (top to bottom are the random functions sampled from a Mat\'{e}rn $3/2$ covariance function, squared-exponential covariance function, and from the combination of both.) In red are ten models for the DIB profile.}
\label{fig:kernels_fit}
\end{figure}

\subsubsection{Initial conditions and the MCMC fit}

Our spectra are populated by many absorption lines and in order to fit the right one, we have to set some initial conditions that will constrain the fit.

We fit six parameters of which four describe the DIB and two are used to describe the noise. Initial parameters for the latter are simply the average characteristic widths of the two covariance functions. The initial parameters for the DIB profile are measured from the spectrum by making a preliminary fit of the DIB profile. The same model as in the final fit is used, but noise or uneven continuum is disregarded. The profile is fitted with the Levenberg-Marquard method and the fitted parameters are used as initial parameters for the final Bayesian fit. Initial parameters for each of 128 walkers are perturbed, so that different walkers start at different initial conditions. Initial parameters for the covariance function widths, DIB profile amplitude, width, and position are now defined by a normal distribution around the original value with a standard deviation of 10\% and the asymmetry defined by a normal distribution around the original value with a standard deviation of 0.05. Perturbation of the initial conditions is unnecessary (but harmless) for fitting prominent DIBs, as posteriors from every walker will converge toward the same distribution. For very weak DIBs, however, there is a possibility that a carefully selected random function can fit the DIB better than the given model or vice versa. In this case we want to be sure that a large parameter space is explored by MCMC in order to get the correct posterior distribution.

128 Walkers are progressed for 250 steps and after that the parameters of the best fit are used as the initial parameters for the final chain. 1000 steps are made with 128 walkers and last 750 steps from each walker are used to sample the posterior.

Whenever an interstellar absorption line that is not a DIB is fitted, the asymmetry parameter is set to 0, is not perturbed and is not fitted.

The values of the initial fit are also used to construct a prior. The reason is that the all the initial fits were inspected by eye and provide additional knowledge that should be used in the prior. This is done mostly to prevent the wrong absorption line be fitted when there are several prominent absorption features around the DIB of interest.

\subsubsection{Priors}
We use priors to prevent the fitted parameters to take values that are not physically possible or outside the expected ranges. Such rigours priors as presented below are usually not needed, as strong DIB profiles in regions without any interfering stellar lines are well constrained even if we use flat priors. But as soon a stellar line of a comparable strength is blended with a DIB, priors assure us that the DIB profile is fitted and not the stellar feature. The amount of interference is then only reflected in the measured uncertainties.

The characteristics of the stellar spectrum component is hard to guess, as it changes between different regions of the wavelength range. If we are fitting a DIB close to a Hydrogen line, for example, one very broad feature is expected, so a large value for $l_{se}$ must be used. In a region populated only by narrower metallic lines, the $l_{se}$ will be smaller. The correct value for $l_{se}$ is therefore best to be fitted. The upper limit for $l_{se}$ is not given, while the lower limit is at $l_{se}=0.22\ \mathrm{\AA}$. The limit is somewhat arbitrary, because it only distinguishes the stellar spectrum noise from the correlated noise. All the noise with the characteristic length above this value is considered of stellar origin and all the noise with the characteristic length below this value is considered to be random correlated noise. If a star happens to have features narrower than that in its spectrum, they will be treated as correlated random noise. This also limits the correlation length for the correlated noise to just under 3 pixels, which makes sense. The prior for the $l_{m3/2}$ therefore has an upper limit of 0.22~{\AA}. It also has a lower limit of 0.08~{\AA}, which is the pixel size. Because the DIBs widths fall into the same range as the $l_{se}$, we make the prior for $l_{se}$ lower at values that are smaller than the FWHM of the DIB.

We first fitted DIBs in most spectra without using any priors in order to get the average profile shapes. The average values are then used to construct priors for the remaining four parameters. Prior for the asymmetry is constructed from the average value only, priors for the FWHM and central wavelength are constructed from the averaged values and the initial fit, and prior for the EW is constructed solely from the initial fit.

Prior for the asymmetry is simply the average asymmetry measured for each DIB individually. Prior has a Gaussian shape with the FWHM of 0.15 and is limited into the range between -0.75 and 0.75.

Prior for the EW is a Gaussian centred on the EW we get from the initial fit. It has a lower limit of EW=0 and an upper limit which is three times the initial EW. Shape of the prior is such that the $\ln P$ drops to 0.1 at the upper limit. Because we visually inspected all initial fits, we are confident that the EW of three times the fitted value is impossible. 

Prior for the FWHM is constructed from the averaged value for the FWHM of each DIB and the value from the initial fit. 

Prior for the central wavelength is a wide peak centred at the average value and a narrower peak centred at the value from the initial fit. The central wavelength is also limited to $\pm$2.5~{\AA} from the average value. Keep in mind that the average central wavelength does not take the selection function of our program stars into the account. 

\begin{figure}
\includegraphics[width=\columnwidth]{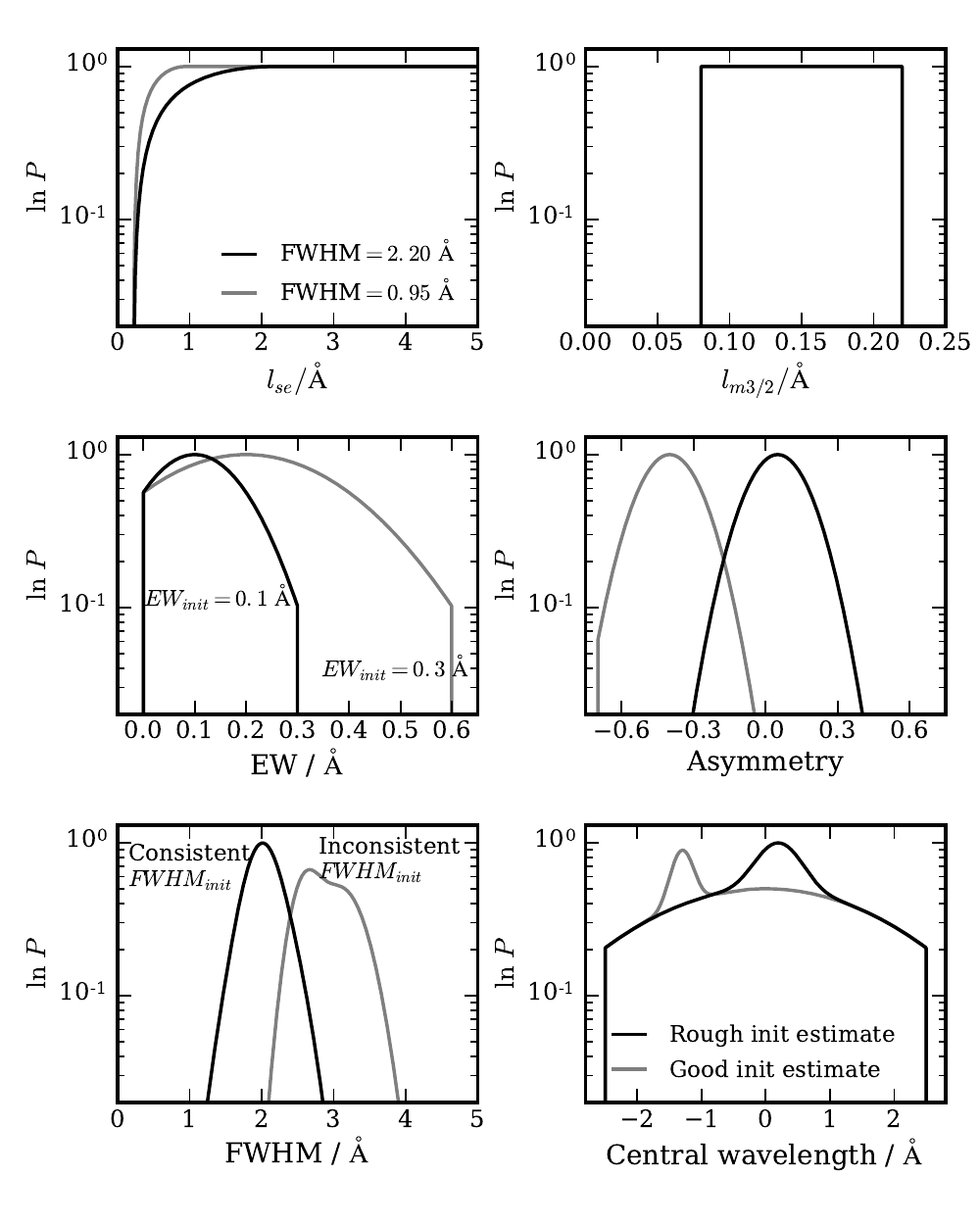}
\caption{Examples of priors for all six fitted variables. Top-left: Prior for $l_{se}$ is flat at larger values but drops to $-\infty$ at $l_{se}=0.22$. At lower values the squared exponential covariance function would fall into the domain of the correlated noise. The prior drops to $-\infty$ gradually, starting at the value of the FWHM of the DIB, preventing the DIB to be fitted by the random function. Top-right: A flat prior is used for $l_{m3/2}$ between 0.08~{\AA} (our sampling and pixel size) and 0.22 (the lower limit for $l_{se}$). Middle-left: Prior for the equivalent width is a Gaussian centred on the initial value. It is scaled as such that the $\ln P$ drops to 0.1 at three times the initial value. We chose the equivalent width to be limited between zero (physical limit) and three times the initial equivalent width (the initial value is precise enough that larger values should be impossible). Middle-right: Asymmetry prior is calculated for each DIB from the distribution of asymmetries we measured without using this prior. A normal distribution is assumed for each DIB but it is limited to $-0.75<asym<0.75$. Bottom-left: Width prior is calculated the same way as the asymmetry prior, except that the initial value is added to the prior. This is not possible for the asymmetry prior, as the asymmetry is less constrained after the initial fit. Two examples are plotted. In black is a prior where calculated and initial widths are similar (FWHM=2.05 and FWHM=2.00) and in gray is a case where the initial width falls into the wing of the calculated distribution (FWHM=3.1 and FWHM=2.6). Bottom-right: Prior for the central wavelength is composed of two components. The wide one is centred at an average value we calculated without using the prior. The narrow component comes from the initial value. The width is proportional to the uncertainty we get from the covariance matrix after the initial fit. Priors after a well constrained (grey) and not so well constrained (black) initial fits are shown.}
\label{fig:priors}
\end{figure}

Examples of a couple priors are shown in Figure \ref{fig:priors}.

\section{Analysis of close pairs}
\label{sec:pairs}

There should always exist LOS that are close enough to each other that the light travels through the same clouds of the ISM. If two LOS also pass the same region of the cloud, we expect the absorption lines produced in the cloud to be the same in both LOS. Assuming some structure in the ISM we will only see differences in LOS when the distance between them approaches or exceeds the size of the structure. 

The two closest LOS we observed are 18 arc seconds apart and the next one 3.9 arc minutes. We cover LOS anywhere between 3.9 arc minutes and 180$^\circ$ without any major gaps. There are 21 pairs of LOSs with the separation less than 30 arc minutes. Assuming a distance of 1~kpc and excluding the difference in distance of the two stars, two LOS with separation of 30 arc minutes will never be more than 8.7 pc apart. 

\subsection{Three models for the structure of the ISM}

\begin{figure}
\centering
Case 1: Different species have different clumpiness
\includegraphics[width=0.92\columnwidth]{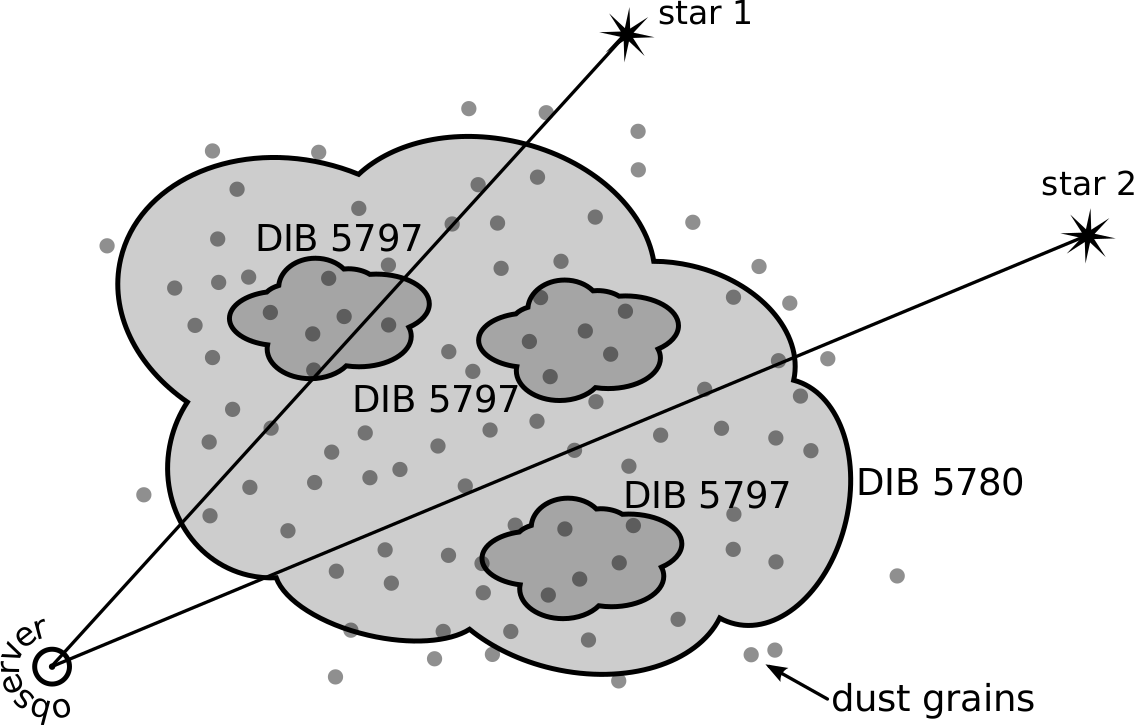}
\caption{Two different species can be in the same cloud of the ISM, but one of them is more clumpy than the other. The difference in the equivalent width between both LOS is therefore more probable to be larger for the more clumpy species. DIBs 5780 and 5797 are used as an example of two different species of the ISM.}
\label{fig:case1}
\end{figure}

\begin{figure}
\centering
Case 2: Different species are in separate clouds
\includegraphics[width=0.92\columnwidth]{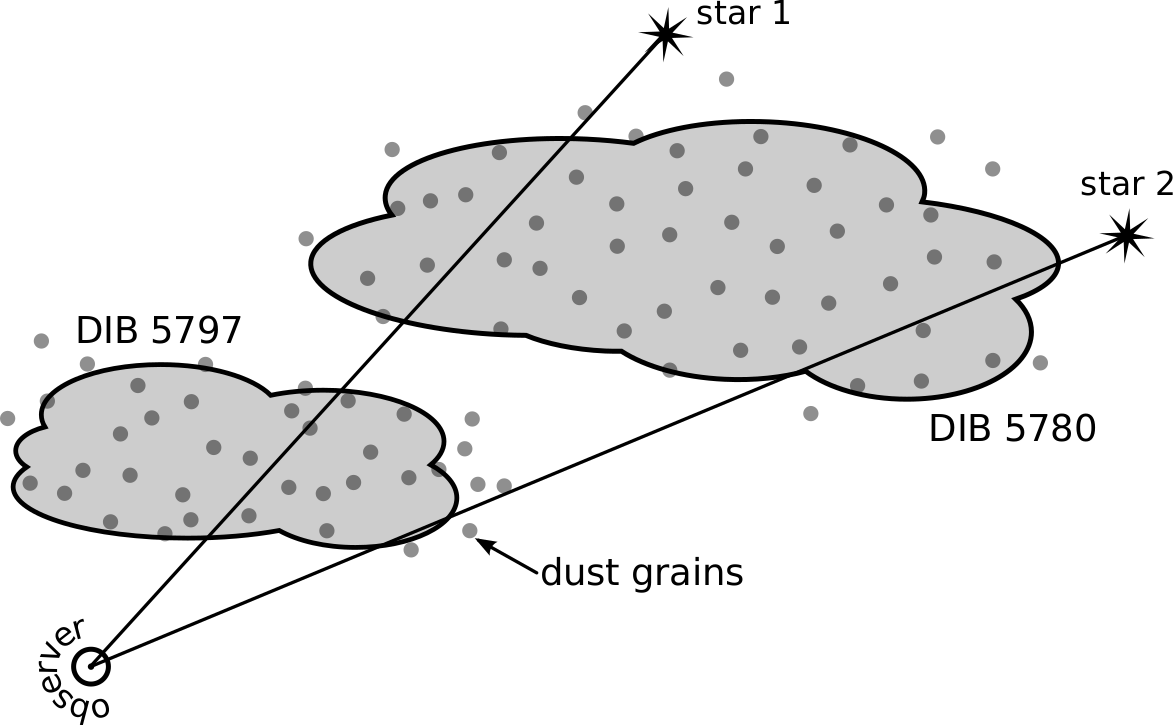}
\caption{There exist two clouds in the LOS. In each cloud one species dominates. }
\label{fig:case2}
\end{figure}

\begin{figure}
\centering
Case 3: Different species create a gradient due to an external influence
\includegraphics[width=0.92\columnwidth]{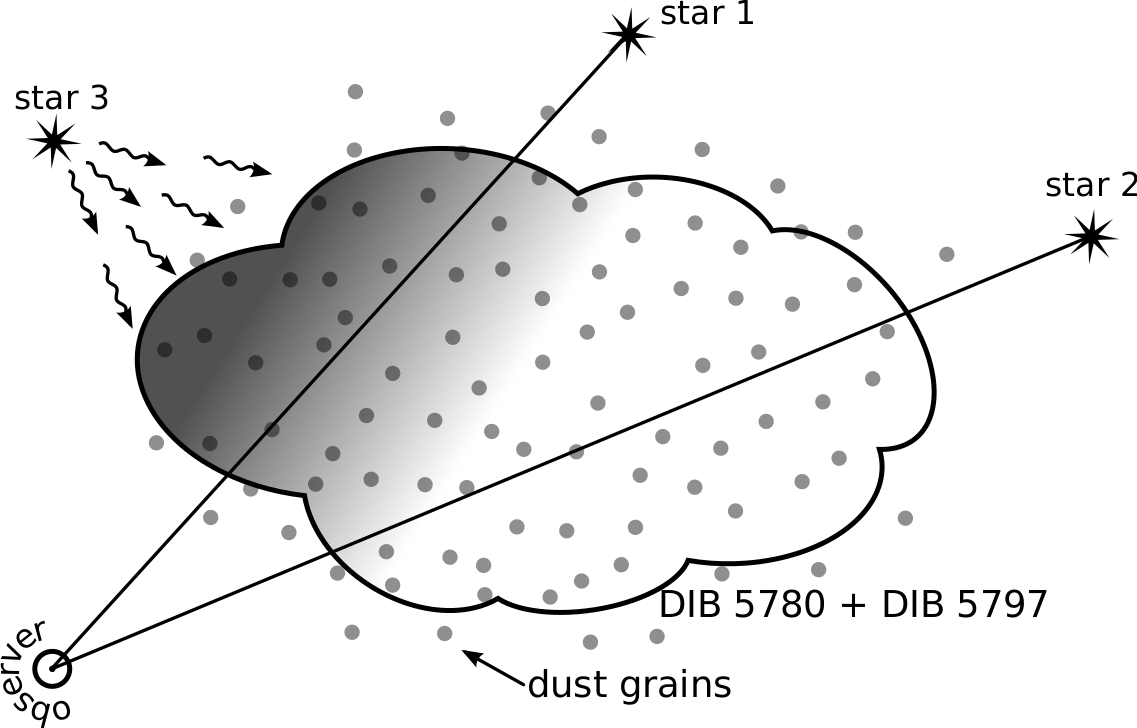}
\caption{A cloud of the ISM in a quiescent state would have similar distribution of both species. If an external source creates a gradient in the physical properties within the cloud, a chemical gradient follows. An external source is not necessarily a star (star 3 in figure), it can be a shockwave, cosmic ray radiation, or other forms of radiation.}
\label{fig:case3}
\end{figure}

There are three possible explanations for differences we observe between DIBs at small scales. First case is one where different DIB carriers are present in the same cloud of the ISM, but the distribution within the cloud is more clumpy for one DIB carrier than the other. Such state is illustrated in Figure \ref{fig:case1}. In the second case different DIB carriers or different groups of DIB carriers exist in two or more clouds, where each carrier or group dominates one cloud. One LOS might therefore pass one cloud but not the other. Such state is illustrated in Figure \ref{fig:case2}. The third option is that there is a gradient within the cloud, so the two LOS pass the same cloud but pass the regions with different composition. A source of the gradient can be a nearby hot star, a shock wave or other external radiation. This case is illustrated in Figure \ref{fig:case3}. This case can be seen as a special case of the first model, but can be tested independently from the first model, so we treat it separately. It is also possible that the actual state of the ISM is a combination of the three options. Our observations are able to detect any combination of the above models as well.

Case 2 can be best confirmed by observing DIBs with different radial velocities. If there exist two separate clouds of gas there is no reason that they should always have the same radial velocity and in as many pairs as we observed, we expect some pairs to have different radial velocities even at small angle-of-separation.

Case 3 can be confirmed if we observe several LOS that sample the same gradient. A smooth transition between the ratios of different DIBs confirms a smooth gradient. If the ratios do not vary smoothly,  the ISM is in the state from case 1. 

\subsection{Illustration of a typical close pair: HD 25638, HD 25639, and HD 25443}

\begin{figure*}
\includegraphics[width=\textwidth]{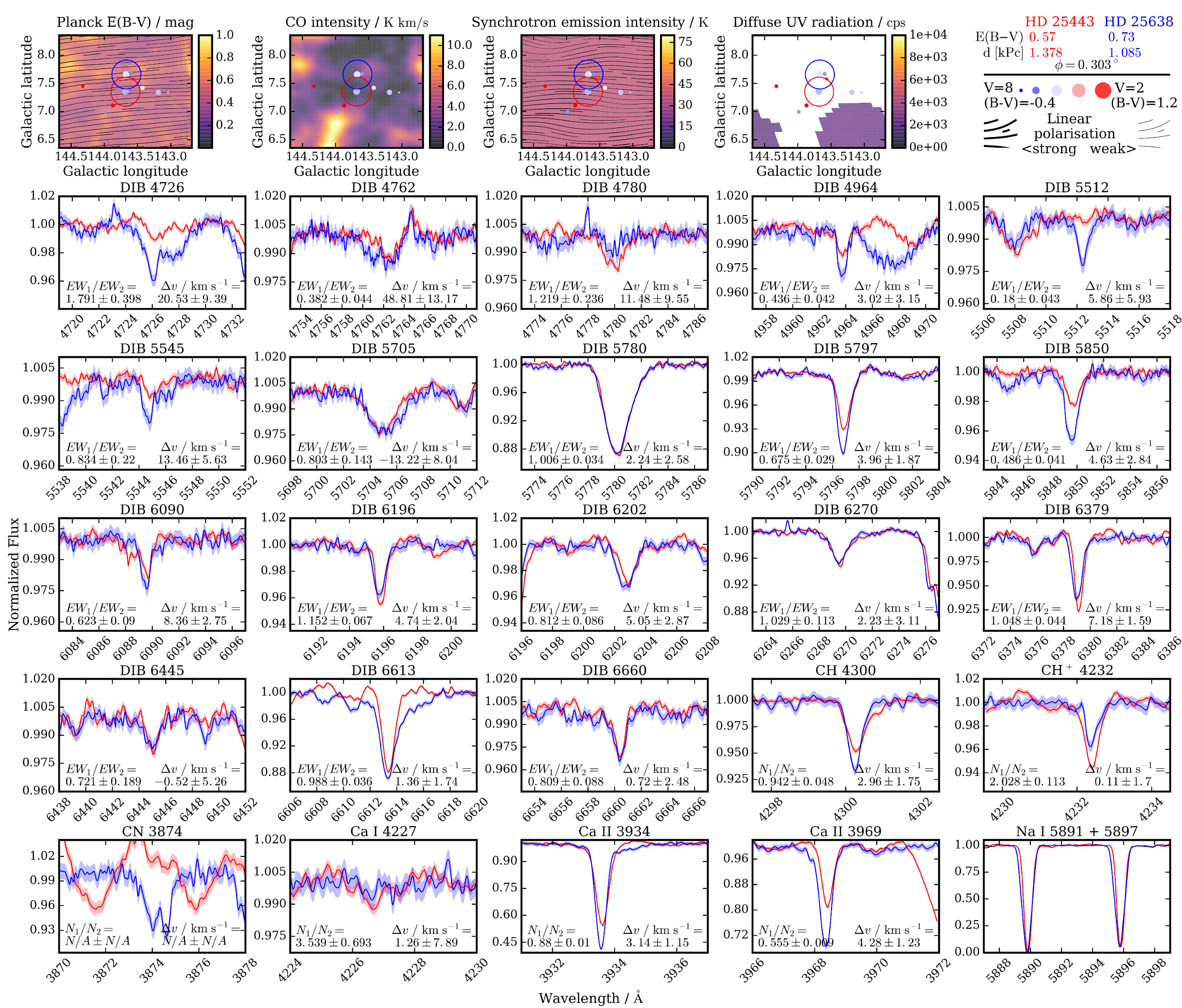}
\caption{Maps of the region around the pair, DIBs, atomic, and molecular lines for HD~25443 (red) and HD~25638 (blue). Some details about the two LOS are given in the top-right corner.} Note the difference in some DIBs.
\label{fig:ppHD25433HD25638}
\end{figure*}

\begin{figure*}
\includegraphics[width=\textwidth]{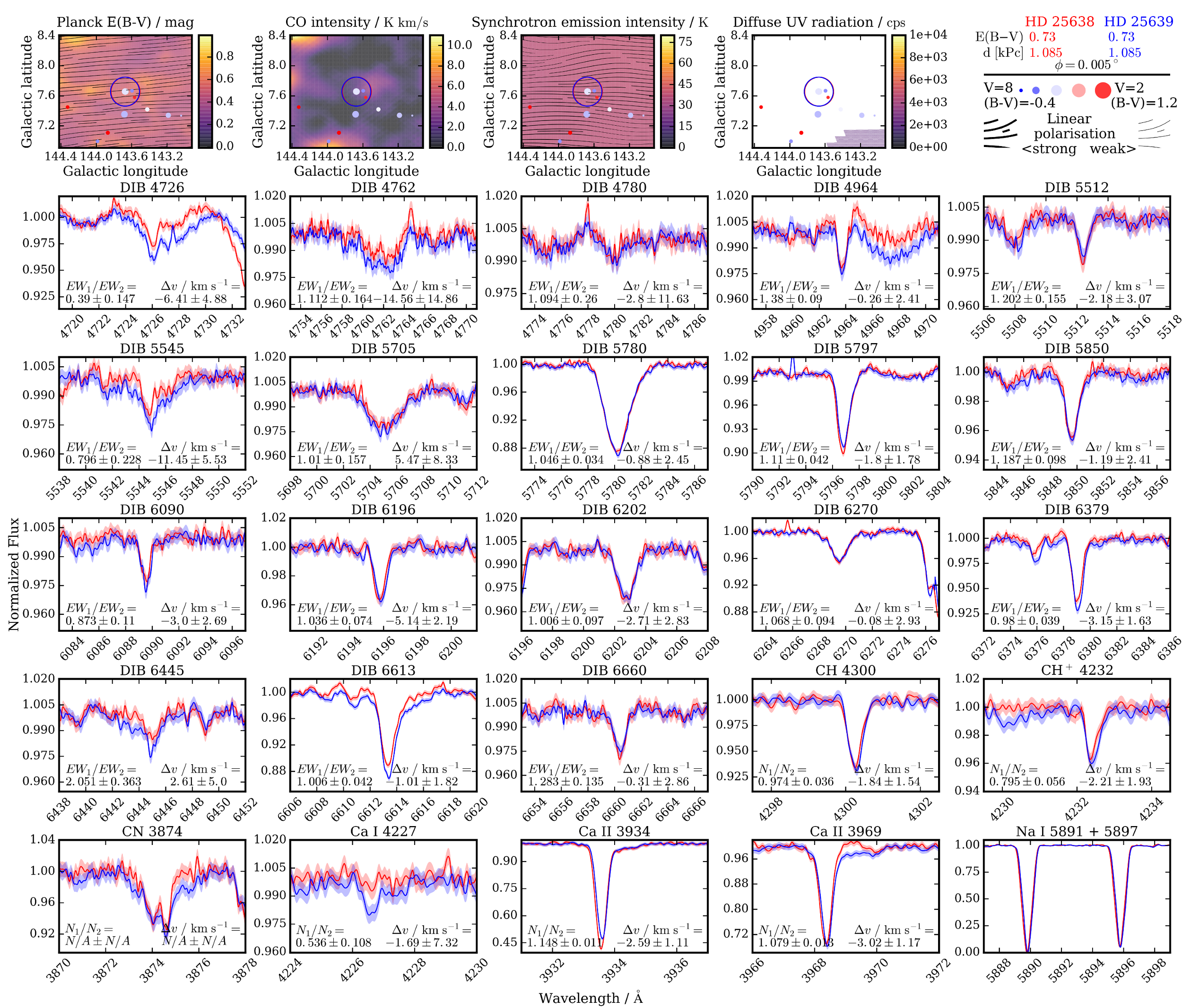}
\caption{Maps of the region around the pair, DIBs, atomic, and molecular lines for HD~25638 (red) and HD~25639 (blue). Some details about the two LOS are given in the top-right corner.} Note that all DIBs are the same in both LOS.
\label{fig:ppHD25638HD25639}
\end{figure*}

An excellent example of a big difference in DIB strengths at small angles-of-separation are pairs HD 25443 and a double star HD 25638 + HD 25639. HD 25443 is 0.3$^\circ$ away from the other two. Figures \ref{fig:ppHD25433HD25638} and \ref{fig:ppHD25638HD25639} show the pairs HD25443, HD25638, and HD25638, HD25639, respectively. Each figure shows all relevant DIBs, atomic and molecular lines. Also shown in each figure is the map with positions of stars on the sky together with dust map and its polarisation, CO emission map, synchrotron emission and its polarisation \citep{planck15}, and near UV diffuse radiation \citep{murthy14}. In the panel showing each absorption line is written the ratio between strengths of both lines in the pair and difference in radial velocity. Figures for other pairs with the angle-of-separation smaller than 1$^\circ$ are available in Appendix \ref{sec:apb}. 

In Figure \ref{fig:ppHD25433HD25638} DIB 5780 is equally strong in both spectra. On contrary, the DIB 5797 is weaker toward HD 25443. The most extreme ratio is seen in DIB 5512, which is almost absent in HD 25443 LOS. DIB 6196 is the only one stronger in HD 25443 LOS. Interesting is also the CH to CH$^+$ ratio being higher in HD 25638/25639 LOS. Note that the color excess of the pair is not the same either. In this case the color excess ratio corresponds to the ratio of the DIB 5797, but this is not the rule in other pairs.

In Figure \ref{fig:ppHD25638HD25639}, where the two LOS are much closer to each other, there are no significant differences between any absorption lines.

If we try to test the three models for the structure of the ISM clouds based on the DIBs observed in the mentioned two pairs, we see that the observations support the first model and do not exclude the third model, but the second model is only plausible if all clouds have the same radial velocity. Figures \ref{fig:case1}, \ref{fig:case2}, and \ref{fig:case3} have been produced with the pair HD25443, HD2563 in mind, trying to explain the observed strengths of DIBs 5780 and 5797.

\subsection{Pairwise correlation}

To describe the differences between EWs between pairs in a more general way, without concentrating on selected pairs like in the previous section, we make a diagram where we plot the difference for every possible pair as it depends on the angle-of-separation.

The difference $d$ between measured EWs of DIBs in LOS $1$ and $2$ is defined as:
\begin{equation}
d=\frac{||EW_1-EW_2||}{EW_1+EW_2}
\end{equation}
or in the case of color excess, the difference is
\begin{equation}
d=\frac{||E(B-V)_1-E(B-V)_2||}{E(B-V)_1+E(B-V)_2}.
\end{equation}

\begin{figure*}
\includegraphics[width=\columnwidth]{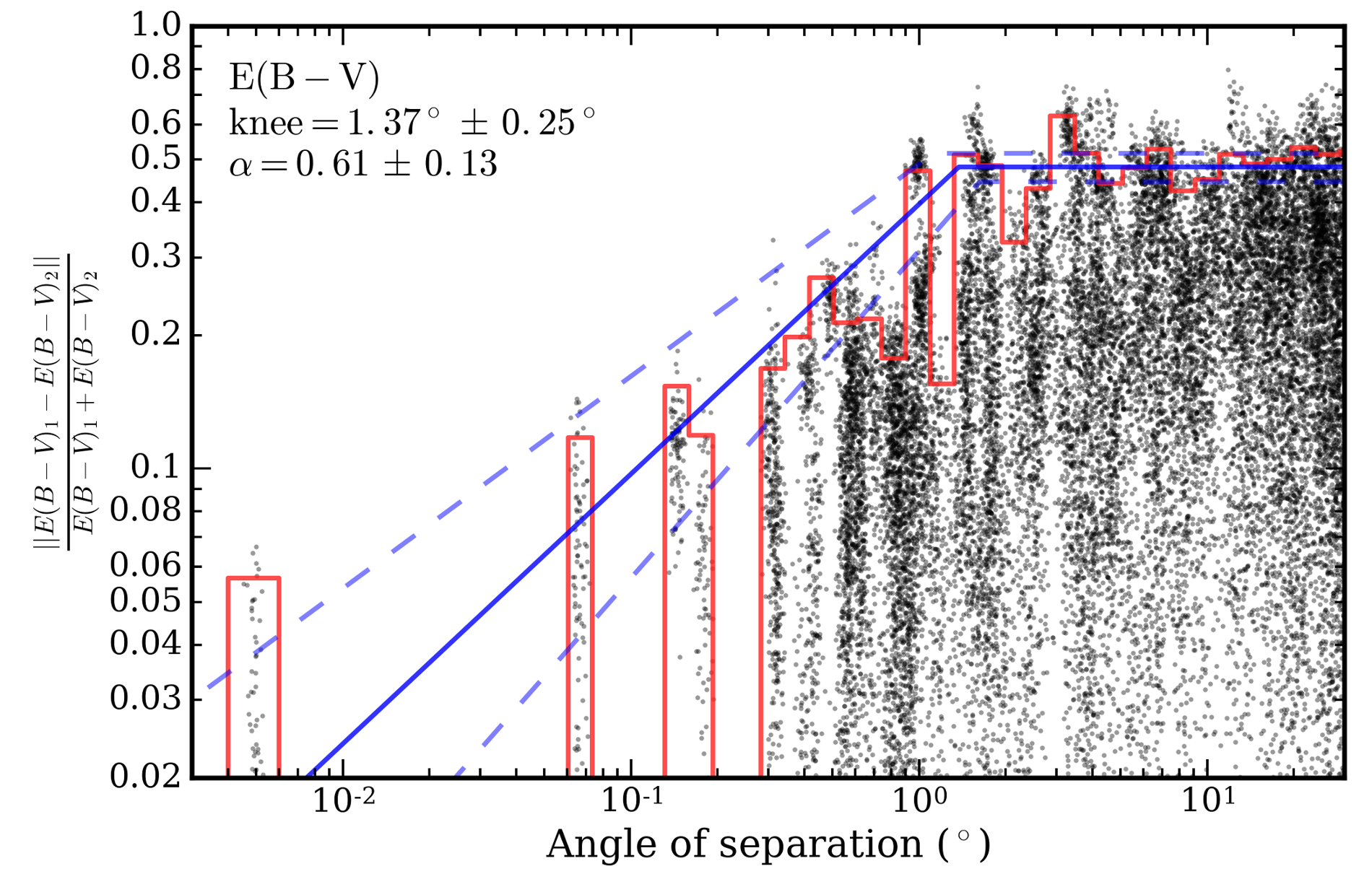} \includegraphics[width=\columnwidth]{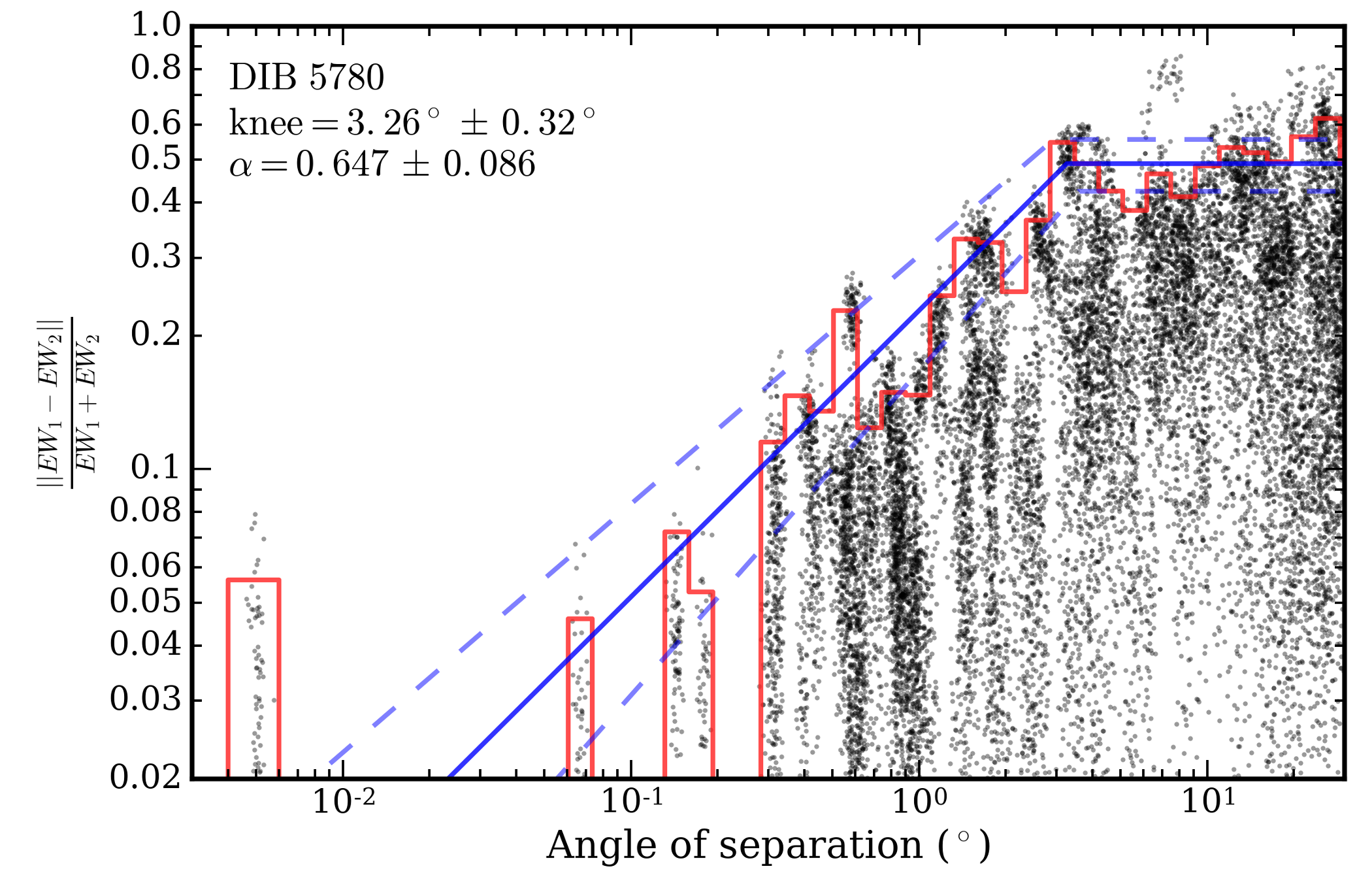}\\
\includegraphics[width=\columnwidth]{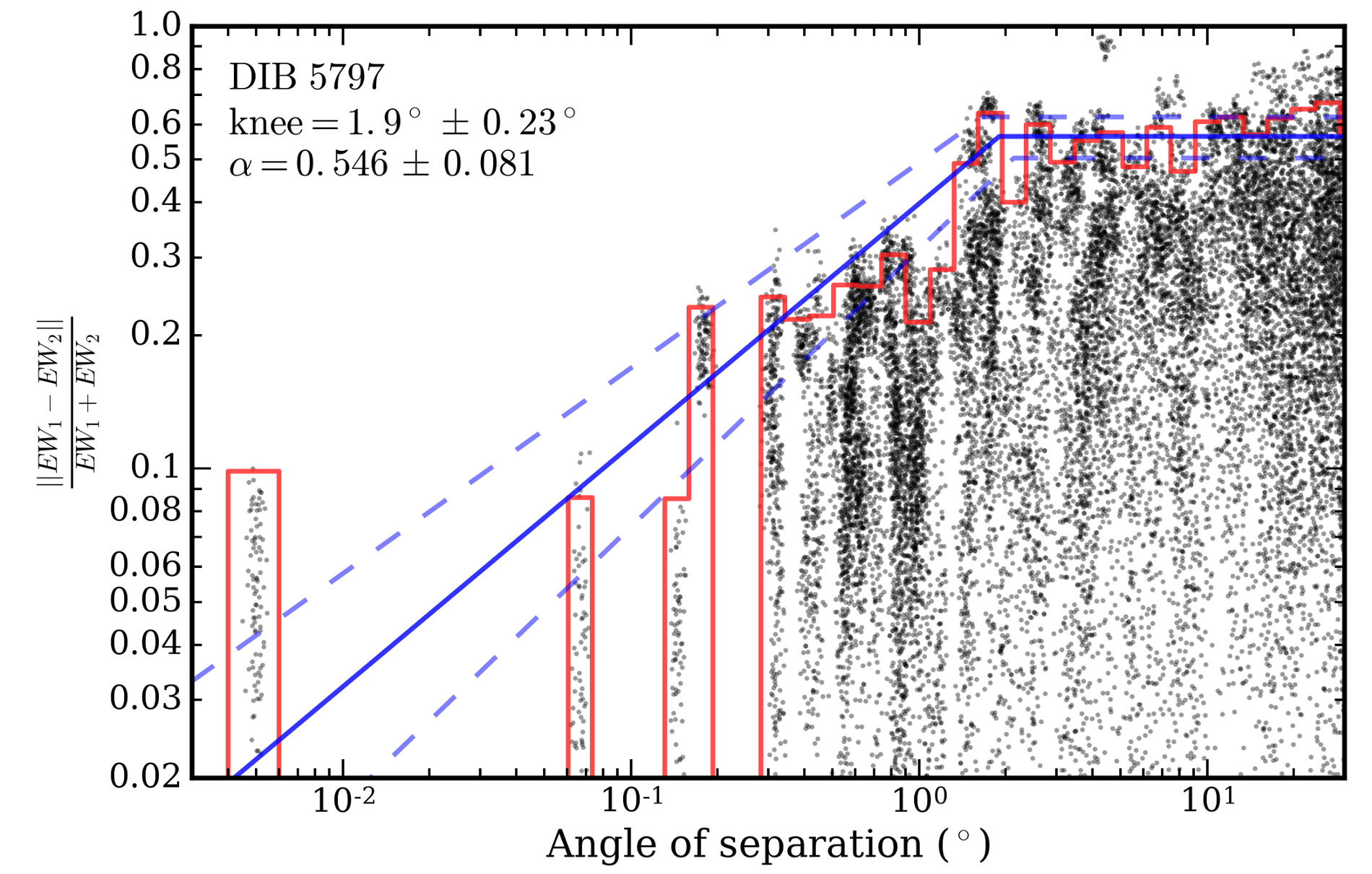} \includegraphics[width=\columnwidth]{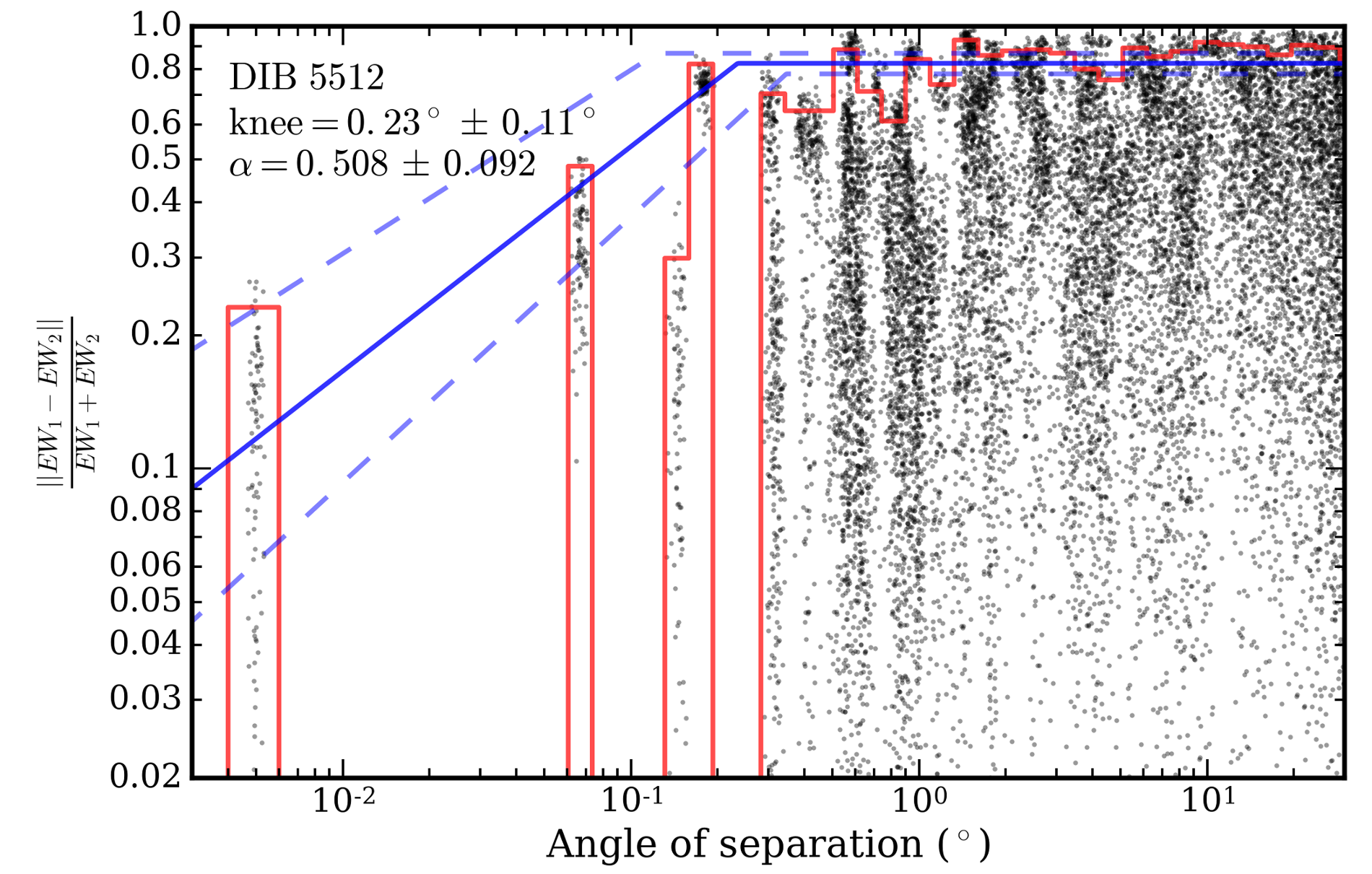}
\caption{Top left: Pairwise correlation between the angle-of-separation and difference in the color excess. To take the uncertainties of the color excess into the account, a normal distribution around the mean value (tabulated in Table \ref{tab:list}) is taken. 100 samples are taken from each distribution which are used to calculate the difference. To plot the distribution, we simply plot the samples. We also scatter the samples around the exact angle separation for the plot to be easier to read. The 95th percentiles of the samples are calculated within log-sized bins (shown in red). A broken power law is fitted and shown in blue. Dashed lines show the uncertainties of the fit. The fitted position of the knee and the slope are given in the top-left corner. Top right: Same as in previous panel, but for the pairwise correlation between the angle-of-separation and difference in the EW of the DIB 5780. The 100 samples are taken from the measured probability density function of the EW. Bottom left: Same as previous panel but for DIB 5797. Bottom right: Same as previous panel but for DIB 5512.}
\label{fig:pwc}
\end{figure*}

The difference is expected to be close to zero at zero angle-of-separation. It is not expected to be exactly zero, as the two stars on the same LOS can be positioned at different distances. It is most likely, however, that the same cloud of the ISM is in the way, so assuming a very small difference is safe. At large separations the expected difference is large, but not necessary one, due to the selection function of our stars. We observe reddened stars, so it is unlikely that any DIB will have a zero EW. More important, the difference is expected to be constant regardless the separation angle, as long as the LOS separation is larger than the angle at which the correlation between different LOS exists. The highest possible difference of one is possible for weaker DIBs where the measured EW is consistent with zero. Thanks to our rigorous approach of estimating the probability density function for each measured EW, such measurements are completely consistent with measurements of stronger DIBs. One can see from Figure \ref{fig:pwc} and Appendix \ref{sec:pwc_a} that a higher maximum difference is indeed more probable for weaker DIBs. Maximum difference is therefore one of the parameters to describe the shape of the pairwise correlation diagram.

In addition to the difference at large angles we introduce two more parameters by fitting a broken power law to the 95th percentiles of the pairwise correlation diagram. The two parameters are the position of a knee (the break in the power law) and the slope at lower angles. In contrast to the difference at large angles, the latter two parameters have an actual physical meaning. The reason we chose the 95th percentile to describe the maximum difference at each angle-of-separation is purely arbitrary. The percentile chosen must be less than 100 to avoid any outliers (statistical or systematic) and choosing any other large percentile gives consistent results. Figure \ref{fig:pwc} shows four pairwise correlation diagrams used to illustrate the main differences between species. The rest of the diagrams are collected in Appendix \ref{sec:pwc_a}. 

The diagrams for three DIBs in Figure \ref{fig:pwc} show that the distribution of DIB 5780 is smoother than that for DIBs 5797 and DIB 5512. In the case of DIB 5780 the knee is at the angle-of-separation of more than 3$^\circ$. The knee is at obviously lower value for DIB 5797, and for DIB 5512 it is practically undetectable as only three data-points lie at angles-of-separation smaller than the position of the knee. In the latter case more data at low angle separation should be taken to actually determine the position of the knee with any significance. The sole data-point at the lowest angle-of-separation is more or less irrelevant to the fitted broken power law, as it has a big uncertainty. The differences at the level of a few percent carry a huge uncertainty, as the measured EWs are only precise to a few percent even for the strongest DIBs. The difference is therefore almost always consistent with zero. It is also consistent with measurements in \citet{cordiner13} made at the same or smaller scale as ours.

\subsection{Constraining the models}
Results presented in the previous section are in strong agreement with the first model, where different DIBs have different clumpiness. 

We were unable to find any tight pairs of LOS where DIBs would have different radial velocities and the angle-of-separation is smaller than the break in the pairwise correlation diagram. We also found no such LOS where different DIBs would have different radial velocities. With the amount of observed LOS it is safe to conclude that the second model, where different DIBs dominate different clouds, can be rejected.

We also did not find any regions with a smooth gradient in a ratio between any two DIBs. We conclude that the gradients only exist at smaller scales then the scales we can reach with our approach or they are completely dominated by the first model.

\begin{table}
\begin{tabular}{lp{1.5cm}p{0.7cm}lp{1.5cm}}
\hline\hline
Line & Position of the knee&& Line &Position of the knee\\
\hline
 & ($^\circ$)&&  &($^\circ$)\\
\hline
DIB 5512 & $0.23\pm0.11$&&DIB 4780 & $0.99\pm0.21$\\

Ca I & $0.53\pm0.21$&&DIB 5850 & $1.05\pm0.29$\\

DIB 5705 & $0.57\pm0.12$&&DIB 6090 & $1.07\pm0.28$\\

DIB 6445 & $0.58\pm0.29$&&Ca II & $1.21\pm0.12$\\

DIB 6379 & $0.64\pm0.04$&&E(B-V) & $1.37\pm0.25$\\

CH & $0.66\pm0.07$&&DIB 6613 & $1.55\pm0.23$\\

CH$^+$ & $0.66\pm0.09$&&DIB 4964 & $1.57\pm0.33$\\

DIB 6270 & $0.70\pm0.12$&&DIB 5797 & $1.90\pm0.23$\\

DIB 4762 & $0.73\pm0.10$&&DIB 6202 & $3.03\pm0.37$\\

DIB 6660 & $0.80\pm0.18$&&DIB 5780 & $3.25\pm0.32$\\

DIB 5545 & $0.95\pm0.22$&&DIB 6196 & $3.55\pm0.53$\\

DIB 4726 & $0.98\pm0.62$&&&\\
\hline
\end{tabular}
\caption{Positions of the knee are collected for all measured DIBs and ISM species in an ascending order.}
\label{tab:results}
\end{table}

\section{Discussion}
\label{sec:disc}

We explored three simple but meaningful models for the spatial distribution of the DIB species. Only one can be confirmed with our data, showing that the DIB species have distinctly different clumpiness. This was showed by plotting pairwise correlations between DIB strengths in single-lined LOS and fitting a two parameter model showing that the correlation breaks at different scales for different DIBs (see table \ref{tab:results} for a condensed list of results). Assuming a typical distance to the observed stars to be 1~kpc, the correlation for DIB 5512 breaks down at 4~pc and for DIB 6196 at 62~pc. Since the actual clouds are closer than the stars, these numbers are only the upper limits. This range is in agreement with scales for some atomic absorption lines calculated in a similar way as ours \citep[e.g.][]{smoker15b}.

We did not explore the second parameter of the fitted model (the slope of the correlation at small scales) as the data is rarely sufficient to constrain it. It can easily be imagined that the slope depends on the fractal structure of the ISM, a relation between big and small clumps of the DIB-bearing medium. More of the closest pairs should have been observed to constrain the slope, but there are not many in the magnitude and color range of the targeted stars. Accuracy of the measured DIB strengths is another concern when measuring the slope, because small differences expected at tight angles-of-separation are hard to measure. More precise measurements require better resolution, higher SNR and ability to remove the stellar absorption lines from the spectra. These are beyond the limitations of our small survey.

Observing finer structure than we targeted in this work is a sensible goal for large spectroscopic surveys where a huge number of close pairs can be observed. This work has even further application to large spectroscopic surveys of stars and the study of the ISM within them. DIBs are commonly used as tracers of the ISM in large spectroscopic surveys. Unlike other ISM absorption lines in the spectra of stars, DIBs are found in all bands covered by contemporary VIS and NIR surveys. Together with the distance to the observed stars they can give a three-dimensional picture of the interstellar medium \citep{kos14,zasowski15}. DIBs also serve as tracers of physical conditions that other absorption lines are not sensitive to \citep{bailey16}. But often the spectra in a small region have to be combined to achieve a SNR required to observe DIBs \citep[e.g.][]{kos14,zasowski15, puspitarini15}. We derived the size of a region in which we can expect the LOS to penetrate the same cloud of the ISM. It turns out that for some DIBs this size is considerably larger than for others. These values (properly modified due to the selection function) can therefore be used in large spectroscopic surveys to amplify the SNR when studying the ISM and select most suitable DIBs for a particular case. The fine structure can then be explicitly measured, instead of inferred like it is in this work.

\begin{figure}
\includegraphics[width=\columnwidth]{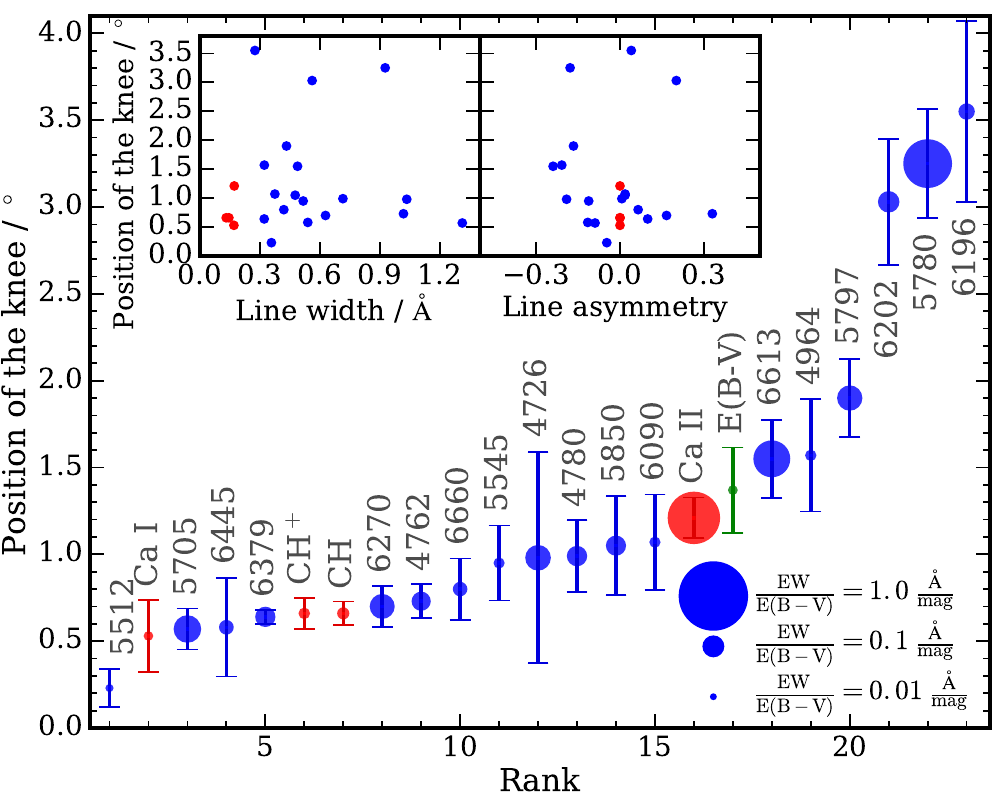}
\caption{Position of the knee for every observed line. Size of the symbol is proportional to an average line strength normalized to color excess. DIBs are plotted in blue and other interstellar lines in red. Two insets in the top-left show the relations between the position of the knee and the remaining two parameters describing the line shape (width and asymmetry).}
\label{fig:result}
\end{figure}

A significant diversity of the clumpiness of the DIB-bearing ISM reveals the nature of the DIB carriers. We expect the DIB carriers with a smoother (less clumpy) distribution to be ionised. Ionised molecules are more likely to be found in the outer layer of the clouds, and are more likely to smooth out the finer structure due to higher temperature. At least DIB~5780 has been attributed to a charged molecule in independens studies \citep{mili14}. However, we can not divide DIB carriers into ionised and neutral molecules solely from the measured clumpiness, as the structure depends on molecular mass and chemistry of the ISM. DIBs 6202, 5780, and 6196 however stand out from the rest (see Figure \ref{fig:result}), as their correlation breaks at angles larger than 3$^\circ$, which is considerably more than for any other DIB.  This suggests that these are absorption lines of ionized molecules. We do not observe any significant correlation between the position of the knee and either strength, wavelength, width, or asymmetry of DIBs.

Pairwise correlation as presented in this paper is a simple description of the shape of the ISM clouds and depends only on a couple parameters related to the physical properties of the actual carriers. It is therefore very convenient for testing the theoretical models for the DIB carriers, should any such models appear in the future. 

\section*{Acknowledgements}
JK is grateful to the Observatory of Padua TAC for a generous amount of observing time and to Maru\v{s}ka \v{Z}erjal, Toma\v{z} Zwitter, and Gregor Traven for assisting with observations. This work is based on observations collected at Copernico telescope (Asiago, Italy) of the INAF - Osservatorio Astronomico di Padova. JK is supported by a Discovery Project grant from the Australian Research Council (DP150104667) awarded to J. Bland-Hawthorn and T. Bedding. JK is grateful to them both for helpful conversations during the course of this research.

\bibliographystyle{mnras}
\bibliography{bib}


\clearpage
\appendix

\newpage

\section{Pairwise correlation diagrams}
\label{sec:pwc_a}

Figures in this section show pairwise correlation between the angle-of-separation and difference in the EW of DIBs. To take the uncertainties of the EW into the account, 100 samples are taken from each measured probability distribution for the EW of the DIB in question. To plot the distribution, we simply plot the samples. We also scatter the samples around the exact angle separation for the plot to be easier to read. The 95th percentiles of the samples are calculated within log-sized bins (shown in red). A broken power law is fitted and plotted in blue (uncertainties are plotted with a dashed line). The fitted position of the knee and the slope are given in the top-left corner of each diagram.

\clearpage

\section{Diagnostic plots for pairs of LOS}
\label{sec:apb}

Plots in this appendix show all the analysed ISM lines in all pairs that are 1$^\circ$ or closer in separation. Each figure shows all relevant DIBs, atomic and molecular lines. Also shown in each figure is the map with positions of stars on the sky together with dust map and its polarisation, CO emission map, synchrotron emission and its polarisation, and near UV diffuse radiation. In the panel showing each absorption line is written the ratio between strengths of both lines in the pair and difference in radial velocity.

\clearpage

\bsp
\label{lastpage}
\end{document}